\documentclass[letterpaper, 11pt]{article}
\usepackage{amsfonts}
\usepackage{amsmath}
\usepackage{amssymb}
\usepackage{enumerate}
\usepackage{natbib}
\usepackage{color}
\usepackage{graphicx, subfigure}
\usepackage{epstopdf}
\usepackage{float}
\usepackage{mathtools}
\usepackage[font=small, margin=1cm]{caption}
\usepackage[left=1in,right=1in,top=1.2in,bottom=1.2in]{geometry}

\usepackage{setspace}
\usepackage{hyperref}
\usepackage[normalem]{ulem} % either use this (simple) or
\usepackage{soul} % use this (many fancier options)
\usepackage{multirow}
\newcommand\E{\ensuremath{\mathbb{E}}}
\newcommand\R{\ensuremath{\mathbb{R}}}

\newcommand\W{\mathcal{W}}

\newcommand{\dx}[1]{ \,d#1 }

\def\R{{\mathbb R}}

\def\P{{\mathbb P}}

\def\1{{\mathbf 1}}

\def\L{{\mathcal L}}

\def\setT{{\mathcal T}}

\def\W{{\mathcal W}}
\def\VV{{\widetilde{V}}}

\newtheorem{theorem}{Theorem}

\newtheorem{corollary}[theorem]{Corollary}

\numberwithin{equation}{section}
\numberwithin{theorem}{section}
\begin{document}
\title{Optimal Risk-Averse Timing of an Asset Sale:\\ Trending vs Mean-Reverting Price Dynamics}
\author{Tim Leung\thanks{IEOR Dept, Columbia University, New York, NY 10027; email:\,\mbox{tl2497@columbia.edu}. Corresponding author.}  \and Zheng Wang\thanks{IEOR Dept, Columbia University, New York, NY 10027; email:\,\mbox{zw2192@columbia.edu}.}}
\date{\today}
\maketitle
\begin{abstract}
This paper studies the optimal risk-averse  timing to sell a risky asset.  The investor's risk preference is described by the exponential, power, or  log utility. Two stochastic models are considered for the asset price --  the geometric Brownian motion  and exponential Ornstein-Uhlenbeck models -- to account for, respectively,  the trending and mean-reverting price dynamics.   In all cases, we derive the optimal thresholds  and certainty equivalents to sell the asset, and compare them  across models and utilities, with emphasis  on their dependence on asset price, risk aversion, and quantity.  We find that the timing option may render the investor's value function and certainty equivalent non-concave in price.  Numerical results are provided to illustrate the investor's strategies and the premium associated with  optimally timing to sell.
\end{abstract}
\vspace{10pt}

\begin{small}
\end{small}

%\tableofcontents
\newpage
\linespread{0.7}

 \section{Introduction}
 We consider  a risk-averse investor who seeks to sell an asset by selecting a timing strategy that  maximizes the expected utility resulting  from the sale. At any point in time, the investor can either decide to sell immediately, or wait for a potentially better   opportunity in the future.      Naturally, the  investor's decision to sell  should depend on the investor's risk aversion and the price evolution of the risky asset.  To better understand their effects, we model the investor's risk preference by  the exponential, power, or log utility.  In addition, we consider  two contrasting models for the asset price --  the geometric Brownian motion (GBM)  and exponential Ornstein-Uhlenbeck (XOU) models -- to account for, respectively,  the trending and mean-reverting price dynamics. The choice of multiple utilities and stochastic models allow for  a comprehensive comparison analysis of all six possible settings.

We analyze  a number of  optimal stopping problems  faced by the investor under different models and utilities.  The investor's value functions and  the corresponding optimal  timing strategies are solved analytically. In particular, we identify the scenarios where the optimal strategies are trivial. These arise in the GBM  model with exponential and power utilities, but not with log utility or under the XOU model with any utility.  The non-trivial optimal timing strategies are shown to be of threshold type. The optimal threshold represents the critical price at which the investor is willing to sell the asset and forgo future sale opportunities. In most cases, the optimal  threshold is determined  from an implicit equation, though under the GBM model with  log utility the optimal threshold is explicit.  Moreover, intuitively the investor's   optimal timing  strategy should depend, not only on risk aversion and price dynamics, but also the  quantity of assets to be sold simultaneously.  In general, we find that the dependence is neither linear nor explicit. Nevertheless, under the GBM model with log utility, the optimal price to sell is inversely proportional to quantity so that the sale will always result in   the same total revenue regardless of quantity. In contrast, under the XOU model with power utility, the optimal threshold is independent of quantity, and thus the total revenue scales linearly with quantity.

While all  utility functions considered herein are concave, the timing option to sell may  render the investor's value functions and certainty equivalents  non-concave in price under different models. For instance, under the GBM model with log utility the value function  can be  convex in the continuation (waiting) region and  concave when the value function coincides with the utility function for sufficiently high asset price. Under the XOU model, we observe that  the value functions are in general neither concave nor convex in price. If the     time of asset sale  is pre-determined and fixed, then the value functions are always concave. Therefore, the  phenomenon of non-concavity  arises due to  the timing option to sell. Mathematically, the reason lies in the fact that the value functions are constructed using  convex functions that are the general solutions to the PDE associated with the underlying GBM or XOU process.  

To  better understand the investor's perceived value of the risky asset  with  the timing option to sell,  we analyze  the certainty equivalent associated with each utility maximization problem.   With analytic formulas, we illustrate the properties of the certainty equivalents. In all cases, the certainty equivalent dominates the current  asset price, and the difference indicates the premium of the timing option. The gap  typically widens as the underlying  price increases before eventually diminishing to zero for sufficiently high price.  As a consequence,  the certainty equivalents are in general neither concave nor convex in price.  If the optimal strategy is trivial, the certainty equivalent is simply  a linear function of asset price.

In  the literature, \cite{Henderson05} considers a risk-averse manager  with a negative  exponential utility who seeks to optimally time the investment  in a project while   trading in a correlated asset as a partial hedge. Under the GBM model, the  manager's optimal timing strategy is to  either   invest according to  a finite  threshold or postpone indefinitely. In comparison, the investor in  our  exponential utility  case  under the GBM model may either sell immediately or at a finite threshold, but  will never find it optimal to wait indefinitely.   \cite{Evans2007a} also study a mixed stochastic control/optimal stopping problem with the objective of determining the optimal time to sell a non-traded asset where the investor has a power utility. In our paper, we show that  the optimal timing  with power utility is either to sell immediately or wait indefinitely under the GBM model, but  the threshold-type strategy is optimal under the XOU model.

 Our study focuses on the GBM and XOU models for the asset price. A related paper by \cite{LeungLiWang2014XOU} analyzes optimal stopping and switching problems under the XOU model. Their results are applicable to our case with   power utility under the same model. Other mean-reverting price  models, such as the OU   model (see e.g. \cite{Ekstrom2011}) and Cox-Ingersoll-Ross (CIR)  model (see e.g. \cite{ewald2010irreversible,LeungLiWang2014CIR}), have been used to analyze various  optimal timing problems.  The recent work by \cite{Ekstrom2016} investigates the optimal timing to sell an asset when its price process follows a GBM-like process with a random drift. All these studies  do not incorporate risk aversion. 
 
 Alternative risk criteria can also be used to study the asset sale timing  problem. Inspired by prospect theory, \cite{henderson2012prospect} considers an S-shape (piecewise power) utility function of  gain/loss relative to the initial price. Under the GBM model,  the investor may find it optimal to sell at a loss. \cite{Pedersen2016} solves for the optimal selling strategy under the mean-variance risk criterion. Instead of maximizing expected utility, one can also incorporate alternative risk penalties to the optimal   timing problems. \cite{LeungShirai} study this problem under both GBM and XOU models with shortfall and quadratic penalties. Other than asset sale, the problem of optimal time to sell and/or buy derivatives by a risk-averse investor has been studied by \cite{henderson2011optimal,LeungLudkovski2}, among others.

 A natural extension for future research is to continue to  investigate the asset sale timing under other underlying  dynamics, such as models with jumps, stochastic volatility, and/or regime switching. Another major direction is to incorporate model ambiguity to the associated  optimal stopping problems; see  \cite{Riedel2009,Cheng2013}. In addition to the trading of risky assets, it is also of interest to incorporate utility and partial hedging  in the optimal liquidation of derivatives.

We organize the rest of the paper in the following manner. In Section \ref{sect-overview}, the asset sale problems are formulated for different utilities and price dynamics.  In Section \ref{sect-results}, we present  the solutions of the problems and discuss the optimal selling strategies. We analyze the  certainty equivalents in Section \ref{sect-ce}. All proofs  are included in Section \ref{sect-proof}.

\section{Problem Overview}\label{sect-overview} We consider a risk-averse asset holder (investor)  with a subjective  probability measure  $\P$.  For our optimal asset sale problems,   we will study two models for the risky asset price, namely, the  geometric Brownian motion (GBM) model and the  exponential Ornstein-Uhlenbeck (XOU) model. First, the GBM price process $S$ satisfies
\begin{align*}
dS_t = \mu S_t\, dt + \sigma S_t \,dB_{t},
\end{align*}with  constant parameters $\mu \in \R$ and $\sigma>0$, 
where $(B_t)_{t\ge 0}$ is a standard Brownian motion under  $\P$. Under the second model, the  XOU  price process   $X$    is defined by
\begin{align} 
X_t &= e^{Z_t}, \notag\\
dZ_{t} &=  \kappa( \theta - Z_t)\,dt+\eta \,dB_{t},\label{OU}
\end{align}
where the log-price   $Z$ is an OU process with  constant parameters $\kappa, \eta>0$, $\theta \in \R$.  

The investor's risk preference is modeled by  three utility functions:
\begin{enumerate}[(1)]
\item Exponential utility
\begin{align*}
U_e(w) = 1 - e^{-\gamma w}, \quad \textrm{for } w \in \R,
\end{align*}
with the  risk aversion parameter $\gamma > 0$;

\item Log utility
\begin{align*}
U_l(w) = \log(w), \quad \textrm{for } w > 0;
\end{align*}

\item Power utility
\begin{align*}
U_p(w) = \frac{w^p}{p}, \quad \textrm{for } w \geq 0,
\end{align*}
where $p := 1- \varrho$, with the risk aversion parameter $\varrho \in [0,1)$. In particular, when $p=1$, the power utility is linear, corresponding to zero risk aversion. 
\end{enumerate}

Denote by $\mathbb{F}$ the filtration generated by the Brownian motion $B$, and $\setT$    the set of all $\mathbb{F}$-stopping times.   The investor seeks to maximize the expected discounted utility from asset sale by selecting the optimal  stopping time.  Denote by $\nu > 0$ the quantity of the risky asset to be sold. For simplicity, we limit our analysis to \emph{simultaneous} liquidation of all $\nu$ units. The investor will receive the utility value of $U_i(\nu S_\tau)$ or  $U_i(\nu X_\tau)$,  $i \in \{e, l, p\}$, under the GBM and XOU model respectively, when all units are sold at time $\tau$. 

Therefore, the investor solves the optimal stopping problems under two different price dynamics:
\begin{align}
&(\textrm{GBM}) \hspace{30pt} V_i(s, \nu) = \sup_{\tau \in \setT}\E_s\!\left\{e^{-r \tau}U_i(\nu S_\tau) \right\},\label{V1a}\\
&(\textrm{XOU}) \hspace{30pt} \VV_i(x, \nu) =  \sup_{\tau \in \setT}\E_x\!\left\{e^{-r \tau}U_i(\nu X_\tau) \right\}, \label{V2}
\end{align} 
for $i \in \{e, l, p\}$,  where $r>0$ is the constant subjective  discount rate. We have used the shorthand notations:  $\E_s\{\cdot\}\equiv\E\{\cdot|S_0=s\}$ and $\E_x\{\cdot\}\equiv\E\{\cdot|X_0=x\}$. By the standard theory of optimal stopping (see e.g. Chapter 1 of \cite{Peskir2006} and Chapter 10 of \cite{Oksendal2003}), the optimal stopping times are of the form 
\begin{align}\label{tauistar}
\tau_i^* &= \inf\{\,t \geq 0: V_i(S_t, \nu) = U_i(\nu S_t)\, \},\\
\widetilde{\tau}_i^* &= \inf\{\,t \geq 0: \VV_i(X_t, \nu) = U_i(\nu X_t)\, \}.\label{tauistar2}
\end{align}

In this paper, we   analytically derive the value functions and show they satisfy their associated variational inequalities. Under the GBM model, for any fixed  $\nu$,  the value functions $V_i(s) \equiv V_i(s,\nu)$, for  $i \in\{e, l, p\}$, satisfy  the variational inequalities 
\begin{align}\label{gbm_variational_inequality}
\textrm{max}\left\{(\L^{S} - r)V_i(s),\, U_i(\nu s) - V_i(s)\right\} = 0, \qquad \forall s \in \R_+,
\end{align}for $i \in\{e, l, p\}$, where  $\L^{S}$ is the infinitesimal generator of  $S$ defined by 
\begin{align}\label{genS}\L^{S} = \frac{\sigma^2s^2}{2}\frac{d^2}{d s^2} + \mu s\frac{d}{d s}.
\end{align}
Likewise, under the XOU model the value functions $\VV_i(x) \equiv \VV_i(x,\nu)$, $i \in\{e, l, p\}$, solve  the variational inequalities 
\begin{align}\label{xou_variational_inequality}
\textrm{max}\{(\L^{Z} - r)\VV_i(e^z)\,,\, U_i(\nu e^z) - \VV_i(e^z)\} = 0, \quad \forall z \in \R,
\end{align}
for $i \in\{e, l, p\}$, where   
\begin{align}\label{genOU}\L^Z = \frac{\eta^2}{2}\frac{d^2}{d z^2} + \kappa(\theta - z)\frac{d}{d z},
\end{align}is the infinitesimal generator of the OU process $Z$ (see \eqref{OU}).
For optimal stopping problems driven by an XOU process, we find it  more convenient to work with the log-price $Z.$

To better understand the value of the risky asset under optimal liquidation, we study the \emph{certainty equivalent} associated with each utility maximization problem. The certainty equivalent is  defined as the guaranteed cash amount that generates the same utility  as the maximal expected utility from optimally timing to sell the risky asset.   Precisely, we define 
\begin{align}
&(\textrm{GBM}) \hspace{-90pt}&C_i(s, \nu) &=  U_i^{-1}\big(V_i(s, \nu)\big), \label{ce_gbm_formulation} \\
&(\textrm{XOU}) \hspace{-90pt} &\widetilde{C}_i(x, \nu)  &= U_i^{-1}\big(\VV_i(x,\nu)\big),  \label{ce_xou_formulation} 
\end{align}
for $i \in \{e, l, p\}$, under the GBM and XOU models respectively. Certainty equivalent gives us a common (cash) unit to compare the values of timing to sell under different utilities, dynamics, and quantities.  

Moreover, the certainty equivalent can shed light on the investor's optimal strategy. Indeed, applying \eqref{ce_gbm_formulation} and \eqref{ce_xou_formulation}  to \eqref{tauistar} and \eqref{tauistar2} respectively, we obtain an alternative expression for the optimal stopping time under each model: 
\begin{align}
\tau_i^* &= \inf\{\,t \geq 0: C_i(S_t, \nu) = \nu S_t \, \},\\
\widetilde{\tau}_i^* &= \inf\{\,t \geq 0: \widetilde{C}_i(X_t, \nu) = \nu X_t\, \}.
\end{align}
In other words, it is optimal for the investor to sell when the certainty equivalent is equals to the total cash amount of $\nu S_t$ or $\nu X_t$, under the GBM or XOU model respectively, received from the sale. 

\section{Optimal Timing Strategies}\label{sect-results}
In this section, we present the analytical results and discuss  the value functions and optimal selling strategies under the GBM and XOU models. The methods of solution and detailed proofs are presented   in Section \ref{sect-proof}. 

\subsection{The GBM Model}\label{sect-results-gbm}
 To prepare for our results for the GBM model, we first consider an increasing general solution to the ODE:
\begin{align} 
\L^{S} f(s)=rf(s), \qquad s\in \R_+,\label{LUGBM}
\end{align}  with $\L^{S}$ defined in \eqref{genS}. This solution is $f(s) = s^{\alpha}$ with
\begin{align}\label{gbm_beta}
\alpha = \left(\frac{1}{2}-\frac{\mu}{\sigma^2}\right) + \sqrt{\left(\frac{\mu}{\sigma^2} - \frac{1}{2}\right)^2 + \frac{2r}{\sigma^2}}\,.
\end{align} 
By inspection, we see that $0 < \alpha < 1$ when $r < \mu$, and $\alpha \geq 1$ when $r \geq \mu.$ 
 
\begin{theorem}\label{thm:GBM_exp_V}
Consider the optimal asset sale  problem \eqref{V1a} under  the GBM model with exponential utility.
\begin{enumerate}[(i)]
\item If $r \geq \mu,$ then it is optimal to sell immediately, and the value function is  $V_e(s, \nu) = 1 - e^{-\gamma \nu s}$.

\item If $r < \mu,$ then the value function   is given by
\begin{align*}
V_e(s,\nu) = \begin{cases}
  ({1 - e^{-\gamma \nu a_e}})(a_e)^{-\alpha} s^{\alpha} &\mbox{if } s \in [0, a_e), \\ 
1 - e^{-\gamma \nu s} & \mbox{if } s  \in [a_e, +\infty),
\end{cases}
\end{align*}
where the optimal threshold $a_e \in (0, +\infty)$ is uniquely determined by  the equation
\begin{align}\label{vgbm_exp_eqn}
\alpha(e^{\gamma \nu a_e} - 1) - \gamma \nu a_e  = 0.
\end{align}
The optimal time to sell  is \[\tau_e^* = \inf\{\,t \geq 0: S_t \geq a_e\, \}.\]
\end{enumerate}
\end{theorem}
\vspace{10pt}
 
Under the GBM model with exponential utility,   the optimal selling strategy can be either trivial or non-trivial. When the subjective discount rate $r$ equals or exceeds the drift $\mu$ of the GBM process, it is optimal to sell immediately. This is intuitive as the asset net discounting  tends to lose value. On the other hand, when $r < \mu$,  the investor should sell   when the unit price   exceeds  a finite threshold. At the optimal sale time $\tau_e^*$, the investor receives the cash amount $\nu a_e$ from the sale of $\nu$ units of $S.$ In other words, $a_e$ is the per-unit  price received upon sale, but according to \eqref{vgbm_exp_eqn} it varies depending  on   the  quantity $\nu$ and   risk aversion parameter $\gamma$.\\

\begin{theorem}\label{thm:GBM_log_V}
Consider the optimal stopping  problem \eqref{V1a} under  the GBM model with log utility. The value function is given by
\begin{align*}
V_l(s,\nu) = \begin{cases}
 \frac{\nu^{\alpha}}{\alpha e}s^{\alpha} &\mbox{if } s \in [0, a_l), \\ 
\log(\nu s) & \mbox{if } s  \in [a_l, +\infty),
\end{cases}
\end{align*}
where $a_l :=  \nu^{-1} \exp({\alpha^{-1}})$  is the unique optimal threshold. The optimal time to sell is \[\tau_l^* = \inf\{t \geq 0: S_t \geq  a_l \}.\]
%\tim{the stopping time is the optimal time to sell, optimal timing. Change for all. }
\end{theorem}
\vspace{10pt}

With  log utility, the optimal strategy is to sell as soon as  the unit price of the risky asset, $S$, enters the upper interval $[a_l, +\infty)$. Note that the optimal threshold $a_l$ is inversely proportional to quantity, so the total cash amount received upon sale,  $\nu a_l =  \exp({\alpha^{-1}})$, remains the same regardless  of quantity. In other words, the log-utility investor is not financially better off by holding more units of $S$.   Under exponential utility, the optimal selling price is implicitly defined by \eqref{vgbm_exp_eqn} in   Theorem \ref{thm:GBM_exp_V} and must be evaluated numerically. In contrast, the optimal threshold under log utility is   fully explicit.  

\begin{figure}[H]
\centering 
\subfigure[]{
\includegraphics[trim = 0.93cm 4 0.93cm 4, clip, width=3in]{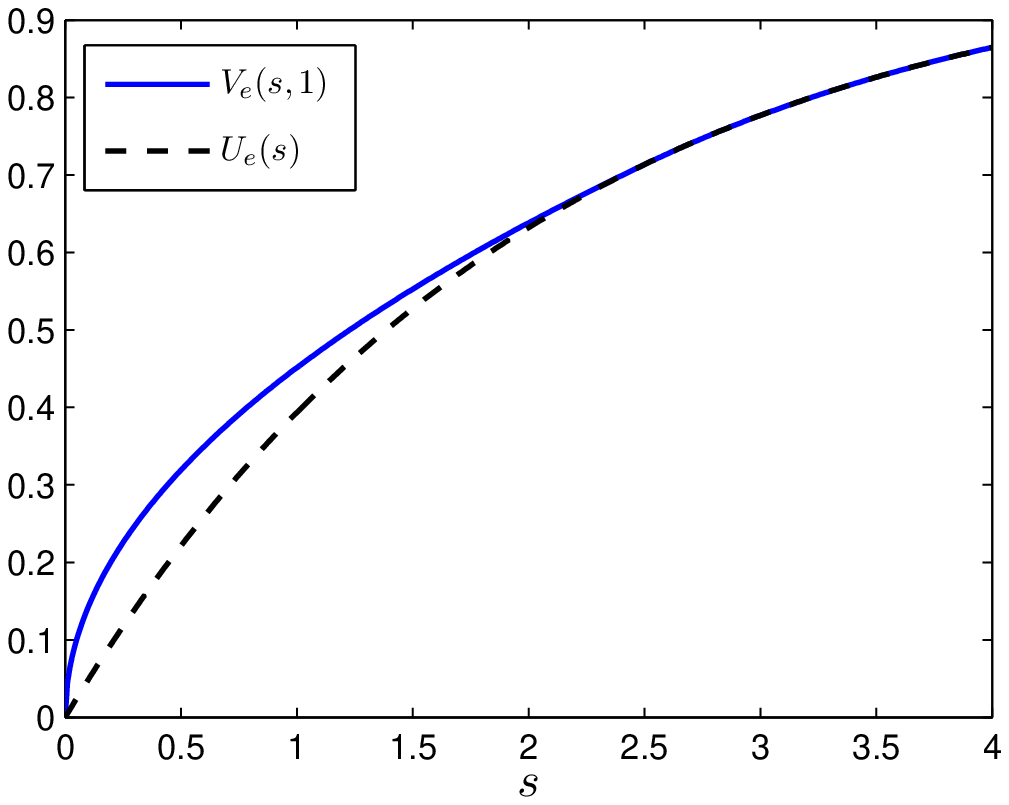}}
\subfigure[]{
\includegraphics[trim = 0.93cm 4 0.93cm 4, clip, width=3in]{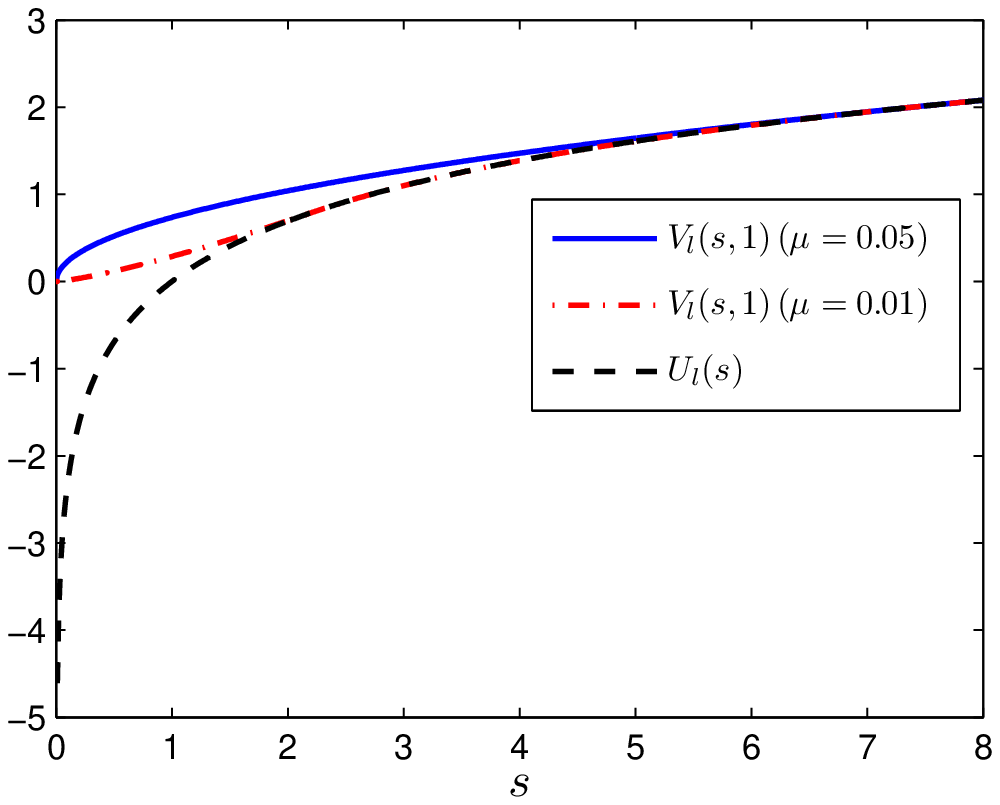}}
\caption{\small{Value functions smooth-paste the utility function under the GBM model. (a) The value function $V_e(s,1)$ dominates the exponential utility $U_e(s)$ (with $\gamma = 0.5$ and $\mu = 0.05$) and coincides for $s \geq a_e = 2.5129$. (b) The value functions  $V_l(s,1)$ (with $\mu = 0.05$) and $V_l(s,1)$ (with $\mu = 0.01$)  dominate the log utility $U_l(s)$ and coincide for $s \geq a_l = 7.3891$ and $s \geq a_l = 2.1832$ respectively. Common parameters: $\sigma = 0.2, \nu=1, r = 2\%.$}}
\label{fig:gbm_pasting}
\end{figure}

Turning to the  value functions, a natural question is whether they preserve the  concavity of the utilities. Indeed, if the investor sells at a pre-determined fixed time $T$, then the expected utility  $W(s) := \E_s\!\left\{e^{-r T}U(\nu S_T) \right\}$ is   concave in $s$ for  any  concave utility function  $U$.  From Theorem \ref{thm:GBM_exp_V}, we see that $V_e(s,\nu)$ is concave in $s$ for all $s \in \R_+.$ On the other hand, $V_l(s,\nu)$ is concave in $s$ when $\alpha < 1$, but it is neither convex nor concave in $s$ when $\alpha \ge 1$.   In other words, the timing option to sell gives rise to the possibility of non-concave value function.    In Figure \ref{fig:gbm_pasting}, we plot  the value functions associated with the  exponential and log utilities when the investor is holding a single unit of the asset. The value functions dominate the utility functions, and coincide smoothly  at the optimal selling thresholds. In Figure \ref{fig:gbm_pasting}(b), the value function  $V_l(s,1)$ under log utility is shown to have two possible shapes. For  $\mu = 0.01< 0.02=r$, i.e. $\alpha < 1$, the value function  $V_l(s,1)$ is   convex when  $s$ is lower than $a_l$, and concave for $s\ge a_l$. In the other scenario, $\mu > r$, i.e. $\alpha >1$, the value function  $V_l(s,1)$ is concave.

 \vspace{10pt}
\begin{figure}[H]
\centering 
\subfigure[]{\includegraphics[trim = 0.93cm 4 0.93cm 4, clip, width=3in]{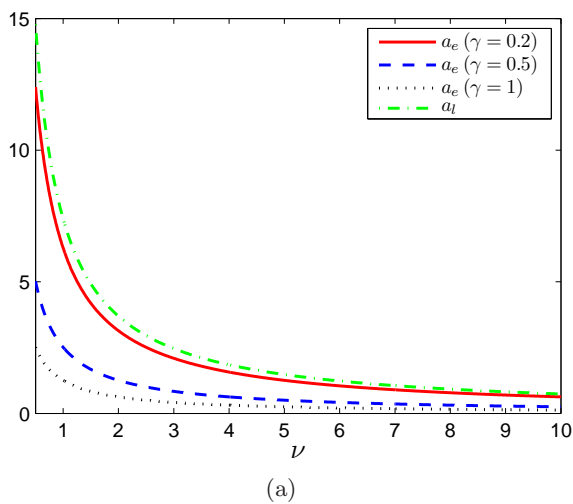}}
%\subfigure[]{\includegraphics[width=3.0in]{xou_price_vs_theta.eps}}
%\subfigure[]{\includegraphics[width=3.0in]{xou_price_vs_eta.eps}}
%\subfigure[]{\includegraphics[trim = 0.93cm 4 0.93cm 4, clip, width=3.0in]{xou_price_vs_n.eps}}
%\subfigure[]{\includegraphics[width=3.0in]{b_vs_p.eps}}
\caption{{\small{Optimal selling thresholds, $a_e$ (with $\gamma = 0.2, 0.5, 1$) and $a_l$, under the GBM model vs quantity $\nu$. Parameters: $\mu = 0.05, \sigma = 0.2, r = 2\%.$}}}
\label{fig:gbm_sensitivity}
\end{figure}

  Figure \ref{fig:gbm_sensitivity} illustrates the effect of quantity $\nu$ on optimal selling thresholds $a_e$ and $a_l$ under exponential and log utilities respectively. The optimal strategy under power utility is trivial and thus omitted from the figure. The optimal threshold $a_e$ is decreasing in $\nu$ for each fixed risk aversion $\gamma = 0.2, 0.5$ and $1$.  Moreover, for any fixed quantity, a higher  $\gamma$ lowers the optimal selling price. The  quantity $\nu$ effectively scales  up the risk aversion to the value  $\nu \gamma$ instead of $\gamma$. Increase in either of these parameters results in higher risk aversion,  inducing the investor to sell at a lower price. In comparison, the log-utility optimal threshold  $a_l$  is explicit and inversely proportional to $\nu$, as seen in the figure.

We conclude this section with a discussion on the optimal liquidation strategy under power utility. First, observe that for any $p \in (0,1],$ the power process $S_t^p$ is also a GBM satisfying  \begin{align*}
dS_t^p = \tilde{\mu}S_t^p\,dt + \tilde{\sigma}S_t^p\, dB_t,
\end{align*}
with new parameters
\begin{align*}
\tilde{\mu} = p\mu + \frac{1}{2}p(p-1)\sigma^2 \quad \textrm{ and } \quad \tilde{\sigma} = p\sigma.
\end{align*} 
Then, the process $\left(e^{-rt}S_t^p\right)_{t\geq0}$ is a submartingale (resp. supermartingale) if $\tilde{\mu} > r$ (resp. $\tilde{\mu} \leq r$). As a result, the optimal timing to sell is trivial, as we summarize next.

\begin{theorem}\label{thm:GBM_pow_V}
Consider the optimal asset sale  problem \eqref{V1a} under  the GBM model with power utility.\begin{enumerate}[(i)]
\item If $\tilde{\mu} \leq r$, then it is optimal to sell immediately, and the value function $V_p(s,\nu) = U_p(\nu s)$.

\item If $\tilde{\mu} > r$, then it is optimal to wait indefinitely, and the value function  $V_p(s,\nu) = +\infty$.
\end{enumerate}
\end{theorem}
\vspace{5pt}

%In this case, the optimal strategy is trivial under all circumstances, i.e. the investor should hold the asset  if the model parameters $\mu$ and $\sigma$, and the risk aversion parameter $p$ are such that $\tilde{\mu} > r$ with the expectation that the discounted utility value of the risky asset will grow indefinitely. Otherwise, it is optimal to liquidate immediately. As a result, $V_p$ is concave in $s$ for all $s \in \R_+.$

\newpage

\subsection{The XOU Model}\label{sect-results-xou}
In this  section, we discuss  the optimal asset sale problems under the XOU model.  As is well known (see p.542 of \cite{borodin2002handbook} and Prop. 2.1 of \cite{alili2005representations}),  the classical  solutions of the ODE\begin{align} \L^Z f(z)=rf(z),\label{LUXOU}\end{align} for $z\in \R$, are \begin{align}
F(z) \equiv F(z;\kappa, \theta, \eta, r):= \int_0^\infty \upsilon^{\frac{r}{\kappa}-1} e^{\sqrt{\frac{2\kappa}{\eta^2}}(z-\theta)\upsilon-\frac{\upsilon^2}{2}} \dx{\upsilon},\label{F}\\
G(z) \equiv G(z;\kappa, \theta, \eta, r):= \int_0^\infty \upsilon^{\frac{r}{\kappa}-1} e^{\sqrt{\frac{2\kappa}{\eta^2}}(\theta-z)\upsilon-\frac{\upsilon^2}{2}} \dx{\upsilon}.\label{G}
\end{align}
Alternatively, the functions $F$ and $G$ can be expressed as
\begin{align*}
F(z) = e^{\frac{\kappa}{2\eta^2}(z - \theta)^2}D_{-r/\kappa}\left(\sqrt{\frac{2\kappa}{\eta^2}}(\theta-z)\right) \quad \textrm{ and } \quad G(z) =  e^{\frac{\kappa}{2\eta^2}(z - \theta)^2}D_{-r/\kappa}\left(\sqrt{\frac{2\kappa}{\eta^2}}(z - \theta)\right),
\end{align*}
where $D_{v}(\cdot)$ is the parabolic cylinder function or Weber function (see \cite{erdelyitranscendental1953}). Direct differentiation yields that   $F'(z) >0$,  $F''(z)>0$,  $G'(z)<0$, and $G''(z)>0$. Hence,  both $F(z)$ and $G(z)$ are strictly positive and convex, and they are, respectively, strictly increasing and decreasing. In particular, the  function $F$ plays a central role in the solution of  the optimal asset sale problems under the XOU model.

\begin{theorem}\label{thm:XOU_exp_V}
Under an XOU model with exponential utility, the optimal asset sale problem admits the solution
\begin{align}\label{XOU_exp_V_sol}
\VV_e(x,\nu ) = \begin{cases}
KF(\log(x)) &\mbox{if } x \in [0, e^{b_e}), \\ 
1 - e^{-\gamma \nu  x} & \mbox{if } x \in [e^{b_e}, +\infty),
\end{cases}
\end{align}
with the constant 
\begin{align*}
K = \frac{1 - \exp\left(-\gamma \nu  e^{b_e}\right)}{F(b_e)} > 0.
\end{align*}
The critical log-price level $b_e \in (-\infty, +\infty)$ satisfies
\begin{align}\label{VXOU_exp_eqn}
\left(1 - \exp\left(-\gamma \nu  e^{b_e}\right)\right)F'(b_e) = \gamma \nu  e^{b_e} \exp\left(-\gamma \nu   e^{b_e}\right)F(b_e).
\end{align}
The optimal time to sell  is 
\begin{align*}
\widetilde{\tau}_e^* = \inf\{t \geq 0: X_t \geq e^{b_e} \}.
\end{align*} 
\end{theorem}

\vspace{10pt}
According to Theorem \ref{thm:XOU_exp_V}, the  investor should sell all $\nu$ units  as soon as the   asset price  $X$ reaches  $e^{b_e}$ or above. The optimal price level $e^{b_e}$  depends on both  the investor's risk aversion and  quantity, but  it stays the same  as long as the product $\nu \gamma$ remains unchanged. 

\begin{theorem}\label{thm:XOU_log_V}
Under an XOU model with log utility, the optimal asset sale problem admits the solution
\begin{align}\label{XOU_log_V_sol}
\VV_l(x,\nu ) = \begin{cases}
DF(\log(x)) &\mbox{if } x \in [0, e^{b_l}), \\ 
\log (\nu x) & \mbox{if } x \in [e^{b_l}, +\infty),
\end{cases}
\end{align}
with  the coefficient
\begin{align}\label{coefD}
D = \frac{b_l + \log(\nu) }{F(b_l)} > 0.
\end{align}
The finite critical log-price level $b_l$ is uniquely determined from the equation
\begin{align}\label{VXOU_log_eqn}
F(b_l) = (b_l + \log(\nu) )F'(b_l).
\end{align}
The optimal time to sell is  
\begin{align*}
\widetilde{\tau}_l^* = \inf\{t \geq 0: X_t \geq e^{b_l} \}.
\end{align*}
\end{theorem}
\vspace{10pt}

\begin{figure}[H]
\centering 
\subfigure[]{
\includegraphics[trim = 0.93cm 4 0.93cm 4, clip, width=3.0in]{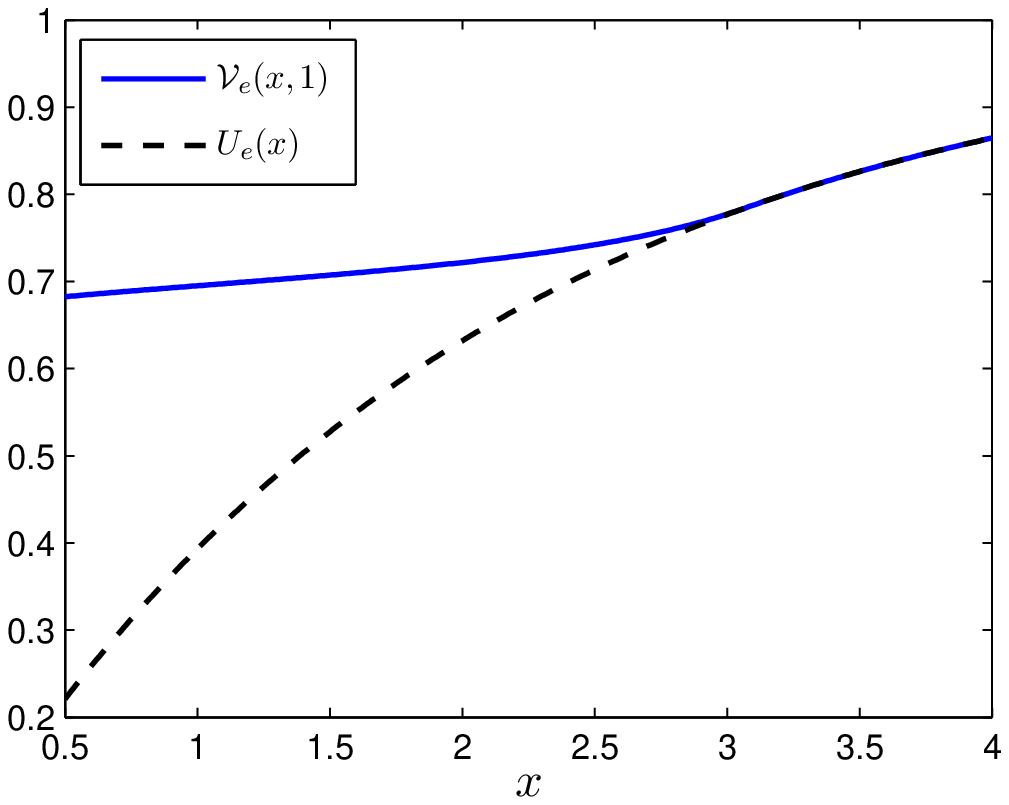}}
\subfigure[]{
\includegraphics[trim = 0.93cm 4 0.93cm 4, clip, width=3.0in]{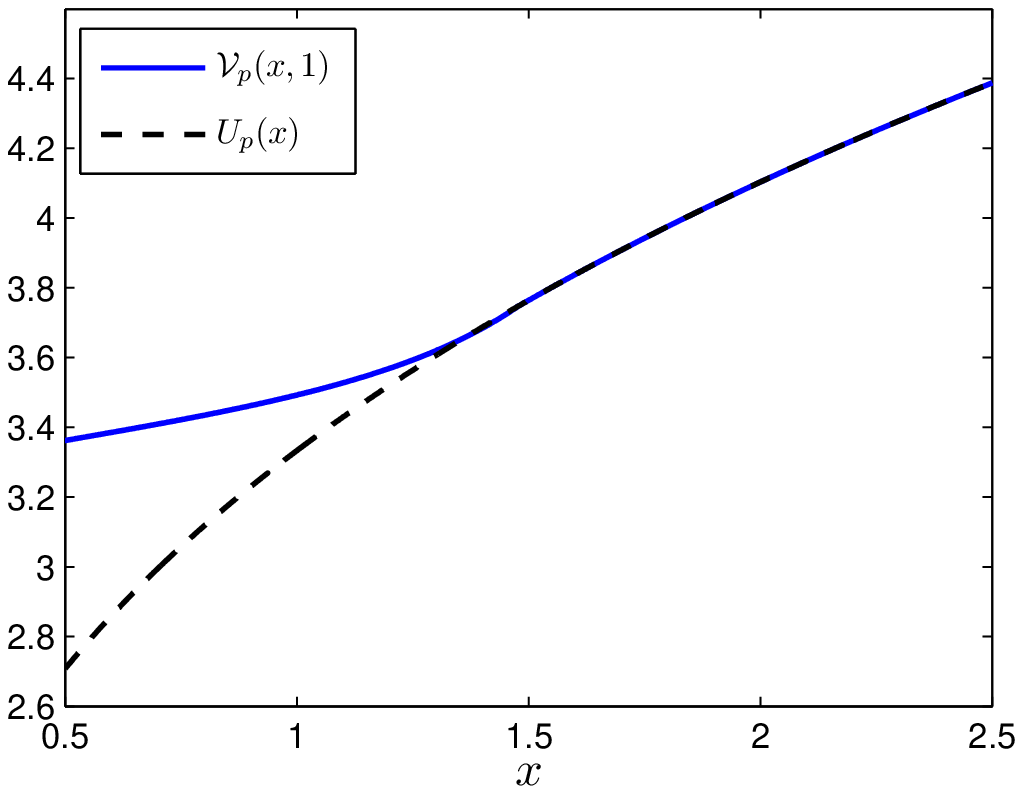}}
\subfigure[]{
\includegraphics[trim = 0.93cm 4 0.93cm 4, clip, width=3.0in]{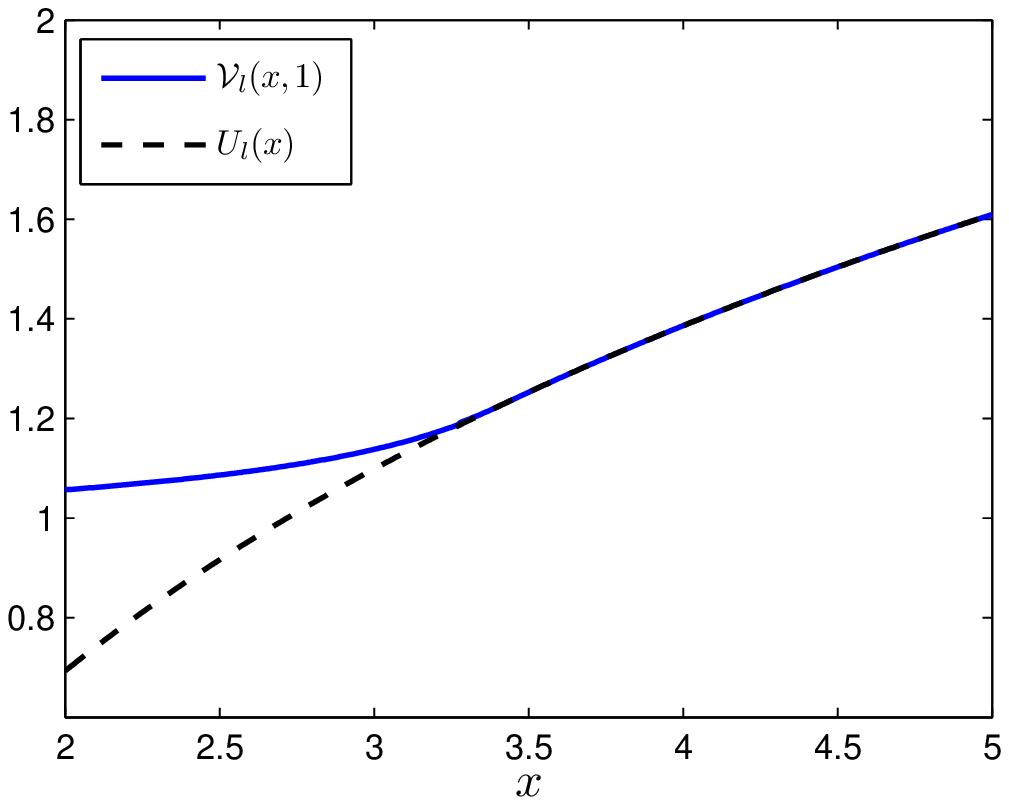}}
\subfigure[]{\includegraphics[trim = 0.93cm 4 0.93cm 4, clip, width=3.0in]{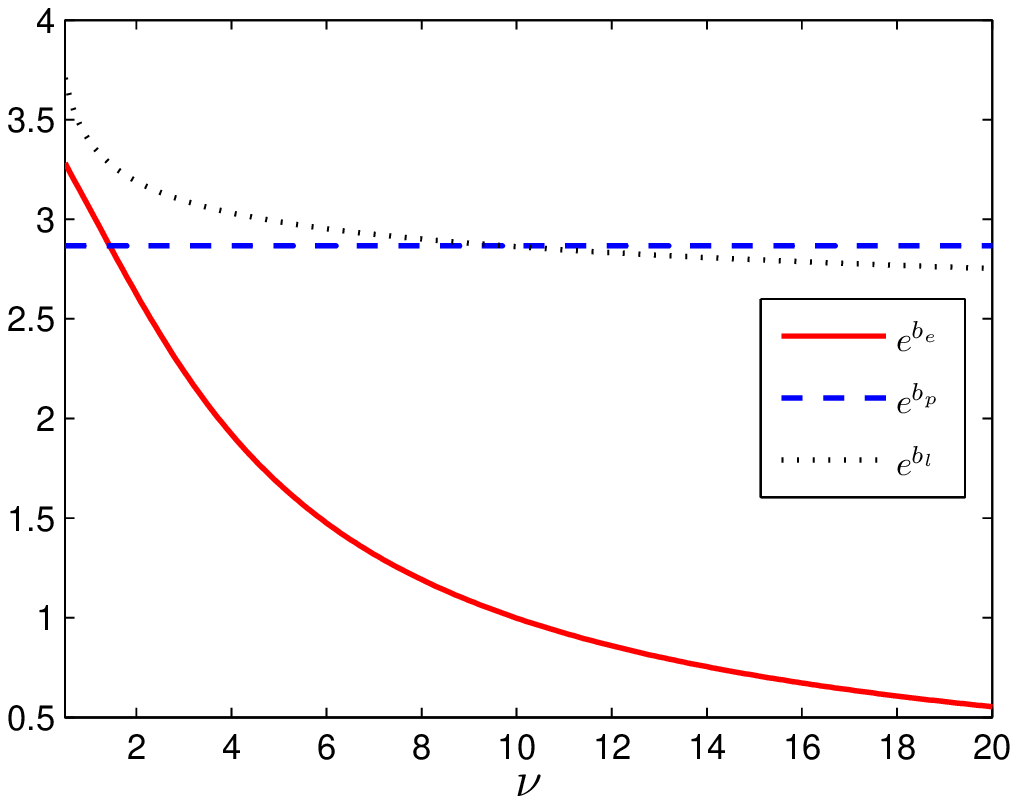}}
\caption{{\small{ Under the XOU model, (a)  the value function $\VV_e(x,1)$ dominates the exponential utility $U_e(x)$ (with $\gamma  = 0.5$) and coincides for $x \geq e^{b_e} = e^{1.1188} = 3.0612.$   (b) The value function $\VV_p(x,1)$ dominates the power utility $U_p(x)$ (with $p=0.3$) and coincides for $x \geq e^{b_p}= e^{0.3519} = 1.3715.$   (c) The value function $\VV_l(x,1)$ dominates the log utility $U_l(x)$ and coincides for $x \geq e^{b_l} = e^{1.2227} = 3.3963.$  (d) Optimal selling thresholds vs quantity $\nu.$ $\gamma = 0.5, p = 0.3.$ Common Parameters: $\kappa = 0.6, \theta = 1, \eta = 0.2, r = 2\%.$ }}}
\label{fig:xou_pasting}
\end{figure}

In Figure \ref{fig:xou_pasting}(d), we see that the optimal unit selling price $e^{b_l}$ is decreasing in $\nu$ but when multiplied by the quantity $\nu$, the total cash amount $\nu e^{b_l}$  received from the sale increases. 

For the case of power utility, we observe that   $\widetilde{X} := X^p$  is again an XOU process, satisfying 
\begin{align}\label{XOU} \widetilde{X}_t = e^{\widetilde{Z}_t}, \quad \text{ where } \quad d\widetilde{Z}_t =  \kappa(\tilde{\theta} - \widetilde{Z}_t)\,dt+\tilde{\eta} \,dB_{t}, \quad t\ge 0,
\end{align}
with the new parameters   $\tilde{\eta} := p{\eta} >0$, and $\tilde{\theta} := p\theta \in \R$. In particular, both the original  long-run mean $\theta$ and volatility parameter $\eta$ have been scaled by a factor of $p$, while the speed of mean reversion remains unchanged.  Therefore, the value function admits the  separable form:
\begin{align}\label{VVp}
\VV_p(x,\nu) &=  \sup_{\tau \in \setT}\E_x\!\left\{e^{-r \tau}\frac{\nu^p X_\tau^p}{p} \right\} = U_p(\nu)\,\VV(\tilde{x},1),
\end{align}
where
\begin{align}
\VV(\tilde{x}, 1) :=  \sup_{\tau \in \setT}\E_{\tilde{x}}\!\left\{e^{-r \tau}\widetilde{X}_\tau \right\}. \label{auxprob}
\end{align}
Hence, without loss of generality,  the optimal timing to sell can be determined from  the optimal stopping  problem  in \eqref{auxprob}, and the corresponding value function $\VV_p$ can be recovered from \eqref{VVp}.

 \begin{theorem}\label{thm:XOU_pow_V}
Under the XOU model with power utility, the solution to the optimal asset sale problem is given by
\begin{align*}
\VV_p(x,\nu ) = \begin{cases}
MF(p\log(x)) &\mbox{if } x \in [0, e^{b_p}), \\ 
\frac{\nu^p}{p}x^p & \mbox{if } x  \in [e^{b_p}, +\infty),
\end{cases}
\end{align*}
where
\begin{align*}
M = \frac{\nu ^p e^{pb_p}}{pF(pb_p)} > 0.
\end{align*}
The critical log-price threshold $b_p \in (-\infty, +\infty)$ satisfies the equation
\begin{align}\label{VXOU_pow_eqn}
 F'(pb_p) = F(pb_p),
\end{align}
where $F(z) \equiv F(z;\kappa,\tilde{\theta}, \tilde{\eta}, r)$. 
The optimal asset sale timing is
\begin{align*}
\widetilde{\tau}_p^* = \inf\{t \geq 0: X_t \geq e^{b_p} \}.
\end{align*} 
\end{theorem}

\vspace{10pt}

The investor should sell all $\nu$ units the first time the asset price reaches the level $e^{b_p}.$  According to  \eqref{VXOU_pow_eqn},  the  optimal price level  is independent of quantity $\nu$, as we can see  in Figure \ref{fig:xou_pasting}(d).

  Under the XOU model, the  value functions $\VV_i(x,\nu), i \in \{e, l, p\}$ are not necessarily concave in $x$ due to the convex nature of $F$ and the   timing option to sell. Let's inspect the value functions in  Figure \ref{fig:xou_pasting}. In each of these three cases,  the value function is  initially convex before smooth-pasting on   the concave utility.

\clearpage
 
In Table \ref{Tab:thresholdsummary}, we summarize the results from Sections \ref{sect-results-gbm} and \ref{sect-results-xou} and list the optimal thresholds for all the cases we have discussed. All thresholds, except  $a_l$, are implicitly determined  by the equations referenced in the table. The  asset price  model plays a crucial role in  the structure of the optimal strategy. Under the GBM model with exponential utility,  immediate liquidation may be optimal in one scenario regardless of the current asset price. On the contrary,  with the same utility under the XOU model, immediate liquidation is never optimal and the investor should wait till the asset price rises to level $e^{b_e}.$ With power utility, the GBM model implies a trivial optimal strategy, whereas the XOU model  results in a threshold-type strategy. Lastly, even though both GBM and XOU price processes lead to non-trivial strategies for log utility, the optimal threshold $a_l$ is explicit while $e^{b_l}$ must be computed numerically.

\renewcommand{\arraystretch}{2}
\begin{table}[H]
\centering
\begin{small}
  \begin{tabular}{c|c|c|c}
        \hline    
	         &      \,\, Exponential utility  & \,\, Log utility \,\, & Power utility\\
	        \hline \hline
	        GBM  & $0\,\, /\,\,a_e$ in \eqref{vgbm_exp_eqn}  & $a_l :=  \frac{e^{1/\alpha}}{\nu}$ &   $0\,\,/\,\,+\infty$\\
	        \hline
	        XOU  & $e^{b_e}$ in \eqref{VXOU_exp_eqn}    & $e^{b_l}$ in \eqref{VXOU_log_eqn}& $e^{b_p}$ in \eqref{VXOU_pow_eqn}\\
        \hline
    \end{tabular}
    \end{small}
 \caption{Optimal  thresholds for asset sale under  different models and utilities.}
     \label{Tab:thresholdsummary}
 \end{table}

\section{Certainty Equivalents}\label{sect-ce}
Having derived the value functions analytically, we now state as corollaries  the  certainty equivalents $C_i(s,\nu)$ and $\widetilde{C}_i(x,\nu), i \in \{e, l, p\}$,  defined respectively in \eqref{ce_gbm_formulation} and \eqref{ce_xou_formulation} under the GBM and XOU models.  Furthermore, to   quantify  the value gained from  waiting to sell the asset compared to   immediate liquidation, we define the \emph{optimal liquidation premium} under each model:
\begin{align}
\text{(GBM)}\quad  L(s, \nu)&:= C_i(s,\nu ) - \nu s,\\
 \text{(XOU)} \quad L(x, \nu)&:= \widetilde{C}_i(x,\nu ) - \nu x,  \end{align} 
for $i \in \{e, l, p\}$. We will examine the dependence of this premium on the asset price and quantity. 

\begin{corollary}
Under the GBM model, the certainty equivalents under different utilities are given as follows:
\begin{enumerate}[(1)]
\item Exponential utility:
\begin{align*}
C_e(s,\nu ) = \begin{cases}
 -\frac{1}{\gamma}\log\left(1 - \frac{1 - e^{-\gamma \nu  a_e}}{{a_e}^{\alpha}}s^{\alpha}\right) &\mbox{if } s \in [0, a_e), \\ 
\nu s & \mbox{if } s  \in [a_e, +\infty).
\end{cases}
\end{align*}

\item Log utility:
\begin{align*}
C_l(s,\nu ) = \begin{cases}
\exp\left(\frac{\nu ^{\alpha}s^{\alpha}}{\alpha e}\right) &\mbox{if } s \in [0, a_l), \\ 
\nu s & \mbox{if } s  \in [a_l , +\infty).
\end{cases}
\end{align*}

\item Power utility:
\begin{align*}
C_p(s,\nu ) = \begin{cases}
\nu s &\, \textrm{ if }\, \tilde{\mu} \leq r,\\
+\infty&\, \textrm{ if }\, \tilde{\mu} > r.
\end{cases}
\end{align*}

\end{enumerate}
\end{corollary}
\vspace{10pt}

With  exponential and log utilities,  the certainty equivalents dominate $\nu s$ --  the value from immediate sale, and they coincide   when the asset price  exceeds the corresponding optimal selling thresholds. With power utility, the investor either sells immediately or waits indefinitely, corresponding to the certainty equivalents of value $\nu s$ and $+\infty$, respectively.

The impact of $\nu$ on  $C_e$ is both direct in its certainty equivalent's expression, but also indirect in the  derivation of  $a_e$. As a result, the relationship between $C_e$ and $\nu$ is rather intricate. In comparison, the  explicit formula for the optimal threshold  $a_l$ under log utility facilitates our analysis on the behavior of the certainty equivalent   $C_l$. Fix any price $s$,  $C_l(s,\nu )$ is   convex in $\nu$  when $r \geq \mu$. Consequently, the liquidation premium is maximized at $\nu = 0.$ However,  when  $r < \mu,$ then  $C_l(s,\nu )$ is concave on the price  interval $(0, \log\left(\frac{1 - \alpha}{s^{\alpha}}\right) + 1)$ and convex on $(\log\left(\frac{1 - \alpha}{s^{\alpha}}\right) + 1,  \exp({\alpha^{-1}})/s).$ This implies that there exists an optimal quantity $\nu^* \in (0, \log\left(\frac{1 - \alpha}{s^{\alpha}}\right) + 1)$ that maximizes  the liquidation premium. This is useful when the investor can also choose the initial position in $S$. \\

Next, we state  the certainty equivalents under the XOU model.

  \begin{corollary}
Under the XOU model,  the certainty equivalents under different utilities are given as follows:
\begin{enumerate}[(1)]
\item Exponential utility:
\begin{align*}
\widetilde{C}_e(x,\nu) = \begin{cases}
 -\frac{1}{\gamma}\log\left[1 - \frac{1 - \exp\left(-\gamma \nu e^{b_e}\right)}{F(b_e)}F(\log(x))\right] &\mbox{if } x \in [0, e^{b_e}), \\ 
\nu x & \mbox{if } x \in [e^{b_e}, +\infty).
\end{cases}
\end{align*}

\item Log utility:
\begin{align*}
\widetilde{C}_l(x,\nu) = \begin{cases}
\exp\left[\frac{b_l + \log(\nu)}{F(b_l)}F(\log(x))\right] &\mbox{if } x \in [0, e^{b_l}), \\ 
\nu x & \mbox{if } x \in [e^{b_l}, +\infty).
\end{cases}
\end{align*}

\item Power utility:
\begin{align}\label{ce_xou_pow}
\widetilde{C}_p(x,\nu) = \begin{cases}
\left[\frac{e^{pb_p}}{F(pb_p)}F(p\log(x))\right]^{1/p}\nu &\mbox{if } x \in [0, e^{b_p}), \\ 
\nu x & \mbox{if } x  \in [e^{b_p}, +\infty).
\end{cases}
\end{align}
\end{enumerate}
\end{corollary}
\vspace{15pt}

For all three utilities, the certainty equivalents  are equal to  the immediate sale value,  $\nu x$,  when  the asset price $x$ is in the exercise  region, where all units are sold.  In addition, we emphasize  that both $b_e$ and $b_l$ are dependent on $\nu,$ which can be observed from \eqref{VXOU_exp_eqn} and \eqref{VXOU_log_eqn}. In contrast, the optimal log-price threshold under power utility $b_p$ is \emph{independent} of $\nu$. Consequently, if we consider any fixed $x$ in the continuation region $(0, e^{b_p}),$ then the certainty equivalent  $\widetilde{C}_p(x,\nu) - \nu x$ is  linear and strictly increasing in $\nu.$ This is interesting since under exponential and log utilities, increasing quantity has the effect of making the investor more risk-averse. In other words, as long as quantity is large enough, the investor will liquidate everything immediately even if the current price appears unattractive.

\begin{figure}[H]
\centering 
\subfigure[]{\includegraphics[trim = 0.93cm 4 0.93cm 4, clip, width=3.0in]{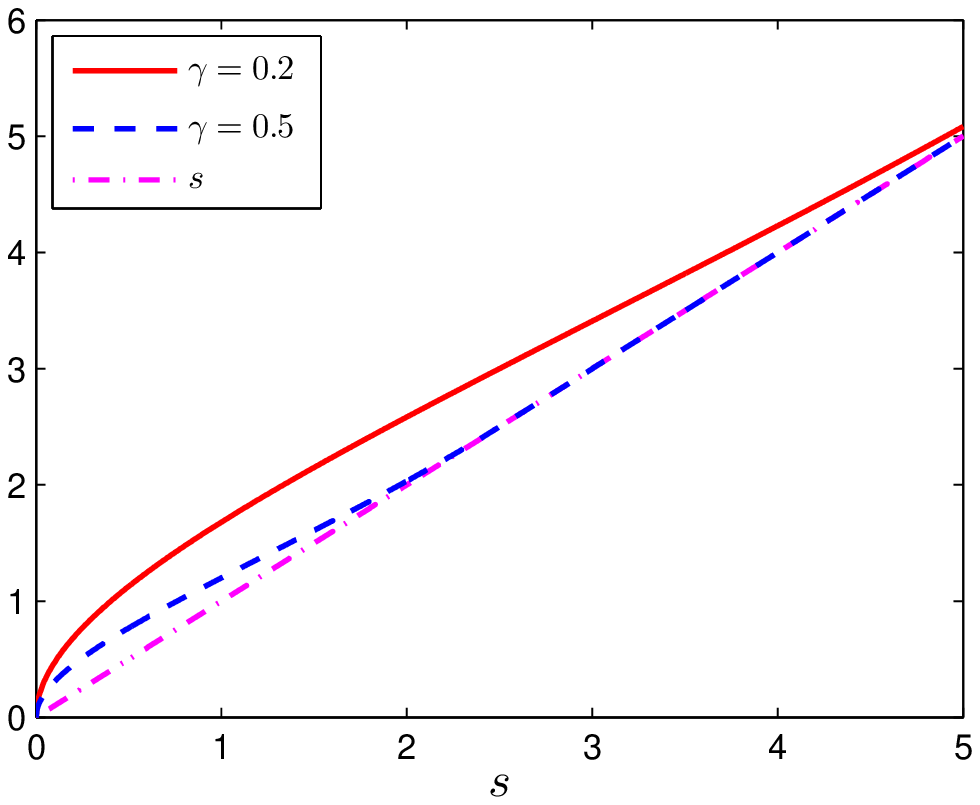}}
\subfigure[]{\includegraphics[trim = 0.93cm 4 0.93cm 4, clip, width=3.0in]{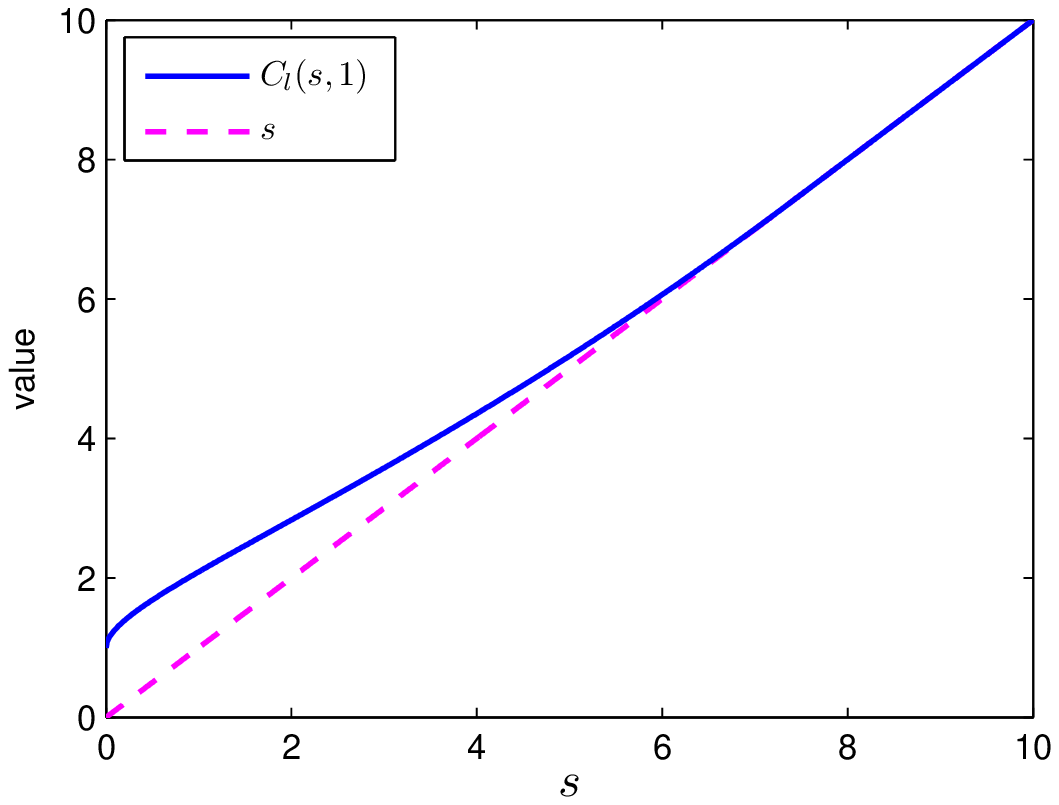}}
\subfigure[]{\includegraphics[trim = 0.93cm 4 0.93cm 4, clip, width=3.0in]{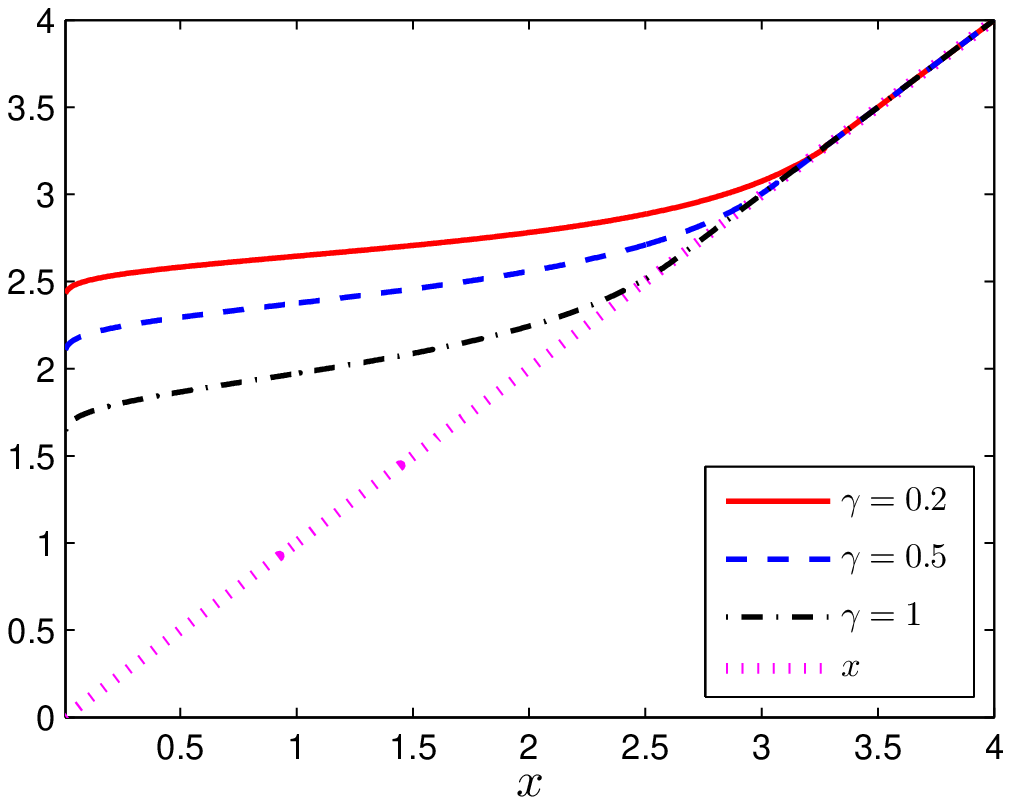}}
\subfigure[]{\includegraphics[trim = 0.93cm 4 0.93cm 4, clip, width=3.0in]{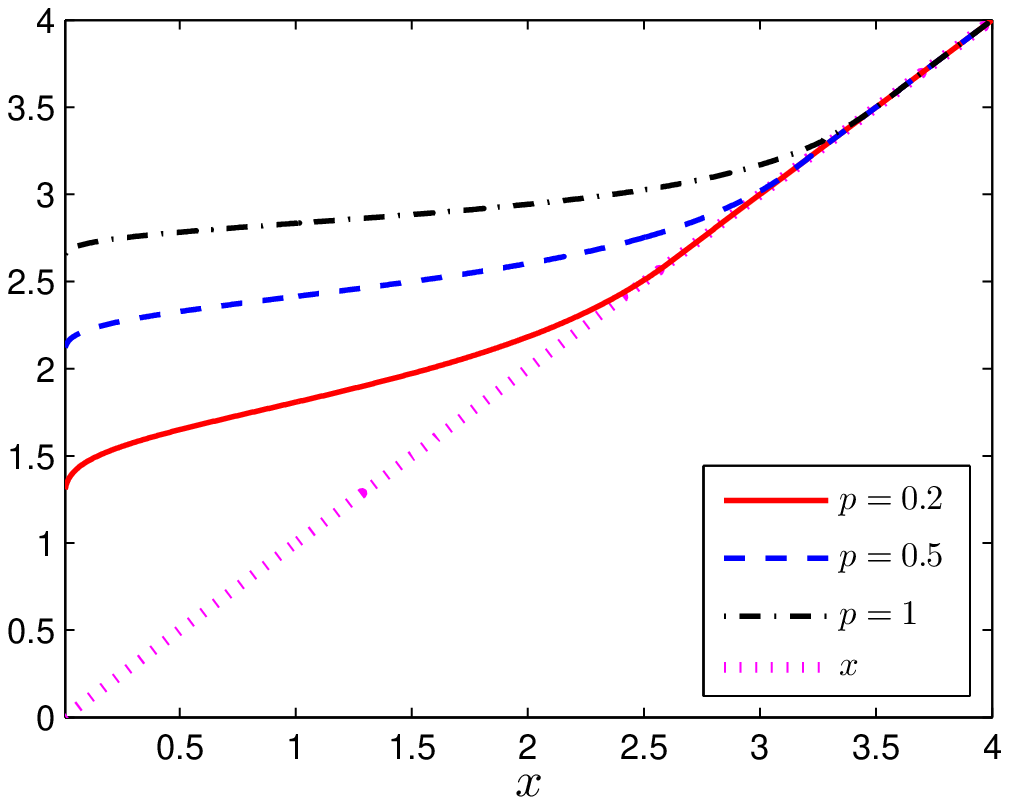}}
\caption{{\small{Certainty equivalent vs price. (a) $C_e(s,1)$ under the GBM model ($\mu = 0.05, \sigma = 0.2$). (b) $C_l(s, 1)$ under the GBM model ($\mu = 0.05, \sigma = 0.2$). (c) $\widetilde{C}_e(x,1)$ under the XOU model ($\kappa = 0.6, \theta = 1, \eta = 0.2$). (d) $\widetilde{C}_p(x,1)$ under the XOU model ($\kappa = 0.6, \theta = 1, \eta = 0.2$). Note that $p:= 1 -\varrho$, where $\varrho$ is the risk aversion parameter. Common parameter: $r = 2\%.$}}}
\label{fig:ce}
\end{figure}

Let us now examine   the certainty equivalents' dependence on the asset price. Under the  GBM model, we plot  the  certainty equivalents, $C_e(s,1)$ and $C_l(s,1),$  against prices, respectively,   in Figures \ref{fig:ce}(a) and   \ref{fig:ce}(b), with a single unit of asset held. The optimal selling strategy for power utility is trivial, and thus, not presented. From Section \ref{sect-results-gbm}, we know that   for sufficiently large $s,$ it is optimal to sell and thus the certainty equivalents will eventually coincide with $s$ and be linear. Notice that in both Figures \ref{fig:ce}(a) and   \ref{fig:ce}(b), the certainty equivalents are concave for small $s$ and subsequently convex for large $s$. In general, the certainty equivalents  are    neither concave nor convex functions of asset price, especially since    the    value functions $V_i$ and $\VV_i,$ $i \in \{e,l,p\}$ are not necessarily concave. 

In Figure \ref{fig:ce}(a), we have also shown  $C_e(s,1)$ for different values of risk aversion level $\gamma$. As    the investor becomes more risk-averse, it becomes  optimal to sell the  asset earlier. This is reflected by  the certainty equivalent's convergence to the linear price line $s$ at a lower price.  Moreover, a less risk-averse  certainty equivalent  dominates a more risk-averse one at all prices. Similar effects of risk aversion is also seen in  Figure \ref{fig:ce}(c) for exponential utility and Figure \ref{fig:ce}(d) for power utility  under the XOU model.

%  Figure \ref{fig:ce}(c) illustrates the behavior of certainty equivalent under exponential utility. We have fixed the quantity at $\nu = 1.$ Since the optimal strategy is defined by a single optimal threshold, the certainty equivalents will converge to $x$ as price increases. Moreover, an investor who is less risk-averse is willing to wait for a higher price. As a result, the certainty equivalent under a larger value of $p$ dominates those with smaller $p$ values.

%Depending on the model, there are a few mechanisms at play. In the case of a GBM price process, the risk-averse nature of the investor prevents further waiting even if the growth rate of the asset is high. Under the XOU model, both the mean-reverting nature of the asset price and the risk-averse utilities discourage the investor from staying in the market.

\begin{figure}[H]
\centering 
\subfigure[]{\includegraphics[trim = 0.9cm 4 0.9cm 4, clip, width=3.0in]{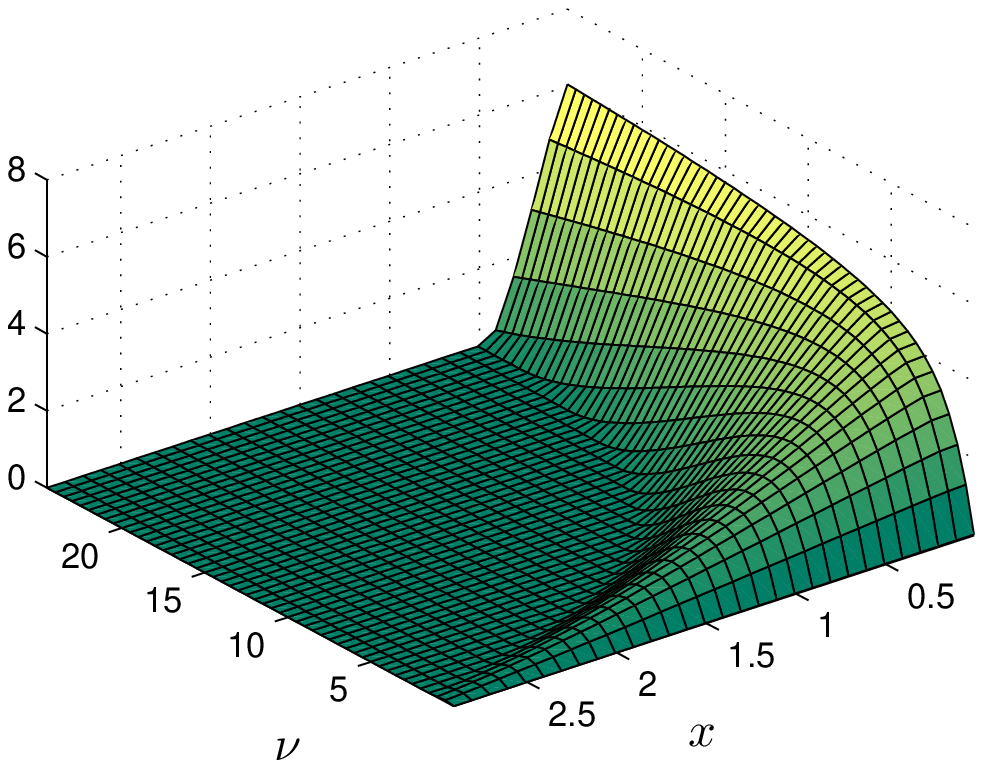}}
\subfigure[]{\includegraphics[trim = 0.9cm 4 0.9cm 4, clip, width=3.0in]{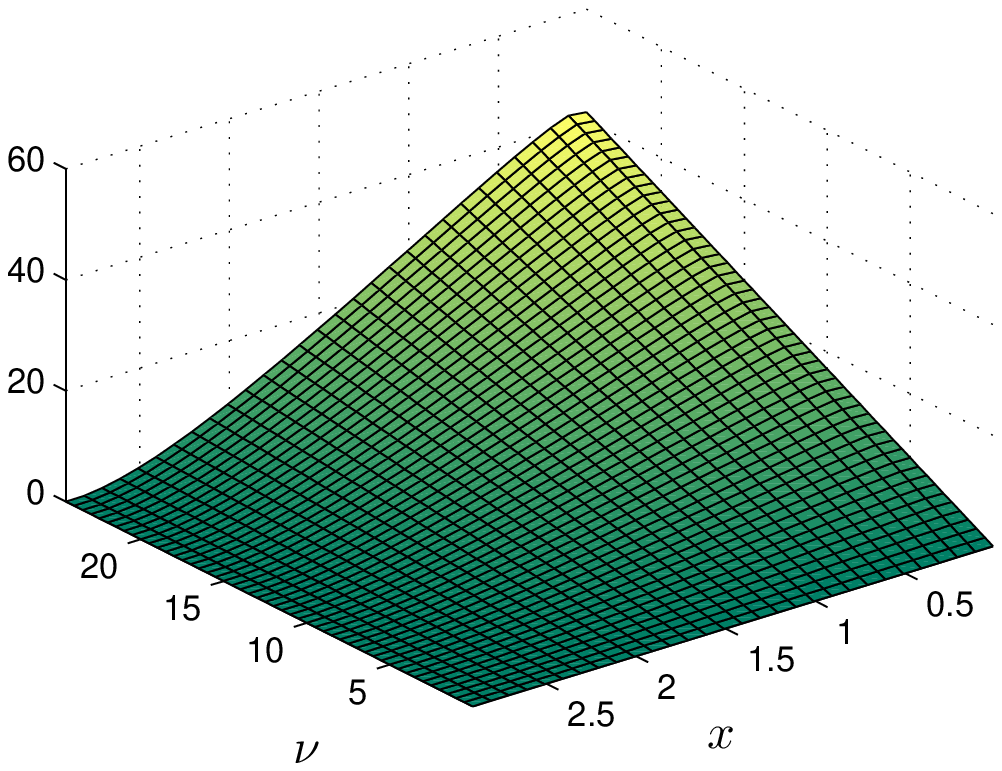}}
\caption{{\small{Liquidation premium $L(x,\nu)$  under the XOU model plotted against  quantity $\nu$ and price $x.$ (a) Exponential utility with $\gamma = 0.5$. (b) Power utility with $p = 0.3$. Common parameters: $\kappa = 0.6, \theta = 1, \eta = 0.2, r = 2\%.$}}}
\label{fig:lp}
\end{figure}

Figures \ref{fig:lp}(a) and \ref{fig:lp}(b), respectively, illustrate the     liquidation premia   for exponential and power utilities under the XOU model.  The liquidation premium for  power utility is linear in $\nu$ for any fixed value of $x$, but the  exponential utility liquidation premium is nonlinear. In general, the liquidation premium  vanishes when  $x$ is sufficiently high when the asset price is in the exercise region.  As we can see from   
Figures \ref{fig:lp}(a) and \ref{fig:lp}(b) and  Figures \ref{fig:ce}(a)-(d), the    optimal liquidation premium tends to be large and may increase when the asset price is very low. This suggests that there is a high value of waiting to sell the asset later if  the current price is low.   As the asset price rises, the premium shrinks to zero. The investor finds no value in waiting any longer, resulting in  an immediate sale.

\section{Methods of Solution and Proofs}\label{sect-proof}
In this section, we present the  detailed proofs for our  analytical results in Section \ref{sect-results},  from   Theorems \ref{thm:GBM_exp_V} - \ref{thm:GBM_log_V}  for  the GBM model to  Theorems \ref{thm:XOU_exp_V} - \ref{thm:XOU_pow_V} for the XOU model. Our  method of solution is  to first construct   candidate solutions using the classical solutions to ODEs \eqref{LUGBM} and \eqref{LUXOU}, corresponding to the GBM and XOU models respectively, and then verify that the candidate solutions indeed  satisfy the associated variational inequalities  \eqref{gbm_variational_inequality} and  \eqref{xou_variational_inequality}.

\subsection{GBM Model}
\paragraph*{Proof of Theorem \ref{thm:GBM_exp_V}}  \textbf{(Exponential Utility).}
 To   prove that the value functions    in Theorem \ref{thm:GBM_exp_V}   satisfy the variational inequality in \eqref{gbm_variational_inequality}, we consider the two cases, 
  $r\geq \mu$ and $r < \mu$, separately. 
  
  When $ r\geq \mu$,     it is optimal to sell immediately. To see this,   for any fixed $\nu$ we verify that $V_e(s,\nu) = 1 - e^{-\gamma \nu s}$ satisfies \eqref{gbm_variational_inequality}. Since $V_e(s,\nu) = U_e(\nu s)$ for all $s \in \R_+,$ we only need to check that the inequality
\begin{align}\label{gbm_exp_VI_trivial}
(\L^{S} - r)(1 - e^{-\gamma \nu s}) = e^{-\gamma \nu s}\left(\mu\gamma \nu s - \frac{\sigma^2\gamma^2 \nu^2 s^2}{2} - re^{\gamma \nu s} + r\right)\leq 0 
\end{align}
holds for all $s \in \R_+$. First, we  observe that the sign of the LHS of \eqref{gbm_exp_VI_trivial}  depends solely on 
\begin{align}\label{gbmexp_g}
g(s) :=  \mu\gamma \nu s - \frac{\sigma^2\gamma^2 \nu^2 s^2}{2} - re^{\gamma \nu s} + r\,.
\end{align}
The first and order second derivatives  $g$ are, respectively, 
\begin{align*}
g' = \mu\gamma \nu- \sigma^2\gamma^2 \nu^2 s - r\gamma \nu e^{\gamma \nu s},  \quad \text{ and } \quad 
g'' = - \sigma^2\gamma^2 \nu^2 - r\gamma^2 \nu^2 e^{\gamma \nu s},
\end{align*}
from which we observe that   $g$ is strictly concave on $\R_+$. Furthermore, $g(0) = 0$ and the fact $\lim_{s \rightarrow +\infty}g(s) = -\infty$ imply that $g$ has a global maximum. Since $g'_+(0) = (\mu - r)\gamma \nu > 0$ (resp. $< 0$) if $\mu > r$ (resp. $\mu < r$), the maximum of $g$ is non-positive if  $r \geq \mu$. As a result, $g$ is non-positive for all $s \in \R_+,$ which  yields  inequality \eqref{gbm_exp_VI_trivial}, as desired.

  For  an arbitrary $\nu >0$, we can  view  $ \nu\gamma$ together as the risk aversion parameter for the exponential utility, and    equivalently consider the asset sale problem with $\nu=1$ and risk aversion $\nu \gamma$  without loss of generality. When $r < \mu$, we consider   a candidate solution $V_e$ of the form $As^{\alpha}$, where $A>0$ is a constant to be determined. Recall that $\alpha$ is less than $1$ when $r < \mu,$ and hence $s^{\alpha}$ is an increasing  concave function. We solve for the optimal threshold $a_e$ and coefficient $A$ from  the value-matching and smooth-pasting conditions
\begin{align}\label{gbm_exp_smooth}
Aa_e^{\alpha} &= U_e(a_e) = 1 - e^{-\gamma \nu a_e}, \\
A\alpha a_e^{\alpha - 1}&= U_e'(a_e) = \gamma \nu e^{-\gamma \nu  a_e}. \notag 
\end{align}
This leads to the following equation satisfied by the optimal threshold $a_e$:
\begin{align}\label{GBM_exp_V_lvl_eqn}
\alpha(e^{\gamma \nu a_e} - 1) - \gamma \nu a_e  = 0. 
\end{align}

We now show that there exists a unique and positive root to \eqref{GBM_exp_V_lvl_eqn}. Our approach involves establishing a relationship between \eqref{GBM_exp_V_lvl_eqn} and $(\L^S-r)(1 - e^{-\gamma \nu s}).$ To this end,  first  observe that the exponential utility $1-e^{-\gamma \nu s}$ has the following properties:
\begin{align}\label{gbmlimit}
\lim_{s \rightarrow 0}\frac{1-e^{-\gamma \nu s}}{s^\beta} = \lim_{s \rightarrow +\infty}\frac{1-e^{-\gamma \nu s}}{s^\alpha} = 0,
\end{align}
where 
\begin{align*}
\beta = \left(\frac{1}{2}-\frac{\mu}{\sigma^2}\right) - \sqrt{\left(\frac{\mu}{\sigma^2} - \frac{1}{2}\right)^2 + \frac{2r}{\sigma^2}}\,, 
\end{align*}
and $s^\beta$ is a decreasing and convex solution to \eqref{LUGBM}. In addition, we have 
\begin{align}
 E_s&\left\{\int_0^{\infty} e^{-rt} \left\lvert (\L^{S}-r)(1-e^{-\gamma \nu s_t})\right\rvert \,dt \right\}\notag \\
&= \E_s\left\{\int_0^{\infty} e^{-rt} \left\lvert e^{-\gamma \nu s_t}\left(\mu\gamma \nu s_t - \frac{\sigma^2\gamma^2 \nu^2 s_t^2}{2} - re^{\gamma \nu s_t} + r\right)\right\rvert \,dt \right\} \notag\\
&< \E_s\left\{\int_0^{\infty} e^{-rt} \left( 1 + \frac{\sigma^2}{2} + r\right) \,dt \right\} = \frac{1}{r} + \frac{\sigma^2}{2r} + 1 < \infty.\label{gbmbound}
\end{align}
The limits in \eqref{gbmlimit} and condition  \eqref{gbmbound} together  imply that the function $1 - e^{-\gamma \nu s}$ admits the following analytic representation:
\begin{align}\label{gbm_exp_analytic}
1 - e^{-\gamma \nu s} = -s^{\beta}\int_{0}^s \Psi^S(\upsilon)(\L^{S}-r)(1-e^{-\gamma \nu \upsilon})\dx{\upsilon} -s^{\alpha}\int_{s}^{+\infty} \Phi^S(\upsilon)(\L^{S}-r)(1-e^{-\gamma \nu \upsilon})\dx{\upsilon},
\end{align}
where
\begin{align*}
\Psi^S(s) = \frac{2s^{\alpha}}{\sigma^2 s^2 \W^S(s)}, \quad \quad \Phi^S(s) = \frac{2s^{\beta}}{\sigma^2 s^2 \W^S(s)}, 
\end{align*}
and
\begin{align*}
\quad \W^S(s) = \frac{2\sqrt{\left(\mu - \frac{\sigma^2}{2}\right)^2 + 2\sigma^2 r}}{\sigma^2} s^{-\frac{2\mu}{\sigma^2}} > 0, \quad \forall s \in \R_+.
\end{align*}
We refer the reader to Section 2 of \cite{zervos2011buy} and Chapter 2 of \cite{borodin2002handbook} for details on the representation. 

Next, dividing  $1 - e^{-\gamma \nu s}$ by $s^{\alpha}$ and differentiating in $s$, we have
\begin{align}\label{gbm_exp_derivative}
\left(\frac{1 - e^{-\gamma \nu s}}{s^{\alpha}}\right)' = \frac{\gamma\nu e^{-\gamma \nu s}s^{\alpha} - \left(1 - e^{-\gamma \nu s}\right)\alpha s^{\alpha-1}}{s^{2\alpha}}.
\end{align}
The crucial step is to recognize that finding the root to the derivative in \eqref{gbm_exp_derivative} is equivalent to solving \eqref{GBM_exp_V_lvl_eqn}. Furthermore, appealing to \eqref{gbm_exp_analytic}, the LHS of \eqref{gbm_exp_derivative} becomes 
\begin{align*}
\left(\frac{1 - e^{-\gamma \nu s}}{s^{\alpha}}\right)' &=  -\left(\frac{s^{\beta}}{s^{\alpha}}\right)'\int_{0}^s \Psi^S(\upsilon)(\L^{S}-r)\left(1 - e^{-\gamma \nu \upsilon}\right)\dx{\upsilon} \\
&\quad \,- \frac{s^{\beta}}{s^{\alpha}}\Psi^S(s)(\L^{S}-r)\left(1 - e^{-\gamma \nu s}\right) + \Phi^S(s)(\L^{S}-r)(1-e^{-\gamma \nu s})\\
&= \frac{\W^S(s)}{s^{2\alpha}}\int_{0}^s \Psi^S(\upsilon)(\L^{S}-r)\left(1 - e^{-\gamma \nu \upsilon}\right)\dx{\upsilon} = \frac{\W^S(s)}{s^{2\alpha}}q_e(s),
\end{align*}
where
\begin{align*}
q_e(s) := \int_{0}^s \Psi^S(\upsilon)(\L^{S}-r)(1-e^{-\gamma \nu \upsilon})\dx{\upsilon}.
\end{align*}
Since both $s^{2\alpha}$ and $\W^S(s)$ are strictly positive for $s > 0$, we conclude that \eqref{GBM_exp_V_lvl_eqn} is equivalent to the equation $q_e(a_e) = 0.$ By differentiating, we obtain
$q_e'(s) = \Psi^S(s)(\L^{S}-r)(1-e^{-\gamma \nu s})$.
Since $\Psi^S(s) > 0$ for all $s \in \R_+$, the sign of $q_e'(s)$ depends solely on $(\L^{S}-r)(1-e^{-\gamma \nu s})$, and thus on $g$ defined in  \eqref{gbmexp_g}.  The   function $g$ is strictly concave on $\R_+.$ Since $g'_+(0) = (\mu - r)\gamma \nu > 0$, the maximum of $g$ is strictly positive. This implies that there exists a unique positive $\varphi$ such that $g(\varphi) = 0.$  
Consequently, we have
\begin{align}\label{gbm_qsign}
q_e'(s)\begin{cases}
> 0 &\, \textrm{ if }\, s <  \varphi,\\
< 0 &\, \textrm{ if }\, s >  \varphi.
\end{cases}.\
\end{align}
This together with the fact that $q_e(0)=0$, lead us to conclude that there exists a unique $a_e>0$ such that $q_e(a_e)=0$ if and only if $\lim_{s\to +\infty}q_e(s) < 0$. The latter holds  due to the facts:\begin{align}\label{gbm_derivative}
q_e(s)   =  \frac{s^{2\alpha}}{\W^S(s)}\left(\frac{1 - e^{-\gamma \nu s}}{s^{\alpha}}\right)', \quad \quad \frac{1 - e^{-\gamma \nu s}}{s^{\alpha}} > 0, \quad \forall s \in [0, +\infty), \quad  \quad \lim_{s \to +\infty}\frac{1 - e^{-\gamma \nu s}}{s^{\alpha}} = 0. 
\end{align}
%By \eqref{gbm_qsign},  $q_e$ is strictly decreasing on $(\varphi, +\infty)$, this means that the statements in \eqref{gbm_derivative} hold true if and only if $\lim_{s\to +\infty}q_e(s) < 0$. 
Therefore, we conclude that there exists a unique finite root $a_e$ to equation \eqref{GBM_exp_V_lvl_eqn}. Furthermore, by the nature of $q_e$ we have
\begin{align}\label{gbm_prop}
a_e>\varphi \quad \textrm{and} \quad q_e(s) > 0, \quad \forall s < a_e.
\end{align}
Finally, from \eqref{gbm_exp_smooth} we deduce that $A = (1 - e^{-\gamma \nu a_e})a_e^{-\alpha} > 0.$ 

Next, we verify the optimality of the candidate solution  using  the variational inequality \eqref{gbm_variational_inequality}. 
First, observe that $1 - e^{-\gamma \nu s} - V_e(s) = 0$ on $[a_e, +\infty)$, and  $V_e(s) \ge 1 - e^{-\gamma \nu s}$ for all $s \in [0,a_e)$.  Lastly, the inequality   
$(\L^{S} - r)(1 - e^{-\gamma \nu s}) \leq 0$  follows from  
\begin{align*}
(\L^{S} - r)(1 - e^{-\gamma \nu s}) &= (\L^{S} - r)As^{\alpha} = 0, \quad \text{ for } s \in [0, a_e),\\
(\L^{S} - r)(1 - e^{-\gamma \nu s})  &\leq 0, \quad \text{ for } s \in [a_e, +\infty).
\end{align*}
Hence,   $V_e(s,\nu)$ given in Theorem \ref{thm:GBM_exp_V} is indeed   optimal.
%Next, on the region $[a_e, +\infty),$
%\begin{align*}
%\end{align*}
%is true due to \eqref{gbm_qsign}. With this, we have shown that $V_e(s,\nu)$ given in Theorem \ref{thm:GBM_exp_V} is indeed the optimal solution to the asset sale problem.
%

% In other words, we require the function $\frac{1 - e^{-\gamma \nu s}}{s^{\alpha}}$ to be increasing on $[0,a_e).$ This is indeed true as a result of \eqref{gbm_derivative} and \eqref{gbm_prop}.

\paragraph*{Proof of Theorem \ref{thm:GBM_log_V}}\label{pf:GBM_log_V} \textbf{(Log Utility).} For any fixed $\nu,$ $\nu S$ follows a GBM process with the same drift and volatility parameters as $S.$ In other words, we can reduce the problem to that of selling a \emph{single} unit of a risky asset whose price process is  $\widetilde{S} := \nu S$ with initial value $\widetilde{S}_0 = \tilde{s} := \nu s.$ Therefore, we construct a candidate solution of the form $V_l(s, \nu)  =  V_l(\tilde{s}, 1)  = B \tilde{s}^{\alpha}$, where $B$ is a positive constant. The value-matching and smooth-pasting conditions are \begin{align}
B\tilde{a}^{\alpha} &= U_l(\tilde{a}) = \log(\tilde{a}), \label{GBM_log_V_smooth_1}\\
B\alpha \tilde{a}^{\alpha - 1}&= U_l'(\tilde{a}) = \frac{1}{\tilde{a}}.  \label{GBM_log_V_smooth_2}
\end{align} 
These equations can be solved explicitly to give a unique solution $\tilde{a} =\exp({\alpha^{-1}})$ and consequently the coefficient $B =  1/{\alpha e}  > 0.$   The optimal aggregate selling price   $\tilde{a}$ translates into the  optimal unit selling price   $a_l = \tilde{a}/\nu = \nu^{-1}\exp({\alpha^{-1}}).$

Next, we  show that $V_l(s, \nu) =  V_l(\tilde{s}, 1) \equiv V_l(\tilde{s})$ given  in Theorem \ref{thm:GBM_log_V} indeed satisfies the variational inequality
\begin{align*}
\textrm{max}\{(\L^{\widetilde{S}} - r)V_l(\tilde{s}), \log(\tilde{s}) - V_l(\tilde{s})\} = 0, \quad   \tilde{s} \in \R_+,
\end{align*}
where $\L^{\widetilde{S}}$ is the infinitesimal generator of the GBM process $\widetilde{S}.$ This is equivalent to showing that $V_l(s,\nu)$ in \eqref{thm:GBM_log_V} satisfies the variational inequality \eqref{gbm_variational_inequality}. First, on $[0, \tilde{a})$ we have $(\L^{\widetilde{S}} - r)V_l(\tilde{s}) = (\L^{\widetilde{S}} - r)B\tilde{s}^{\alpha} = 0.$ Next, observe that the function
\begin{align*}
(\L^{\widetilde{S}} - r)\log(\tilde{s}) &= -\frac{\sigma^2}{2} + \mu - r\log(\tilde{s}),
\end{align*}
has a unique root at $\widetilde{\varphi} = e^{\frac{\mu - \sigma^2/2}{r}}$ such that
\begin{align*}
(\L^{\widetilde{S}} - r)\log(\tilde{s})\begin{cases}
> 0 &\, \textrm{ if }\, \tilde{s} <  \widetilde{\varphi},\\
< 0 &\, \textrm{ if }\, \tilde{s} >  \widetilde{\varphi}.
\end{cases}
\end{align*}
On $[\tilde{a}, +\infty),$ we have $(\L^{\widetilde{S}} - r)V_l(\tilde{s},1)  = (\L^{\widetilde{S}} - r)\log(\tilde{s}).$ To show that  $(\L^{\widetilde{S}} - r)V_l(\tilde{s},1) \leq 0$ we need to prove that $\tilde{a} > \widetilde{\varphi},$ or equivalently,
\begin{align}\label{gbm_log_inequality}
\frac{1}{\alpha} > \frac{\mu - \sigma^2/2}{r}.
\end{align}This follows directly  from the definition of $\alpha$ in \eqref{gbm_beta}. 
%By substituting the value for $\alpha$and rearranging the terms, we can rewrite \eqref{gbm_log_inequality} as 
%\begin{align*}
%\sigma^2r + \left(\mu - \frac{\sigma^2}{2}\right)^2 &> \sqrt{\left(\mu - \frac{\sigma^2}{2}\right)^4 + \left(\mu - \frac{\sigma^2}{2}\right)^22\sigma^2r},
%\end{align*}
%and by squaring both sides, we obtain
%\begin{align*}
%\left(\mu - \frac{\sigma^2}{2}\right)^4 + \left(\mu - \frac{\sigma^2}{2}\right)^22\sigma^2r + \sigma^4r^2 &> \left(\mu - \frac{\sigma^2}{2}\right)^4 + \left(\mu - \frac{\sigma^2}{2}\right)^22\sigma^2r,
%\end{align*}
%which is indeed true.

 Next, we check that $V_l(\tilde{s}, 1)  \ge \log(\tilde{s})$ for all $\tilde{s} \in \R_+$. Since $V_l(\tilde{s},1) = \log(\tilde{s})$ on $[\tilde{a}, +\infty)$, it remains to show  that $\log(\tilde{s}) \leq V_l(\tilde{s},1)$   on $[0, \tilde{a}).$ Using \eqref{GBM_log_V_smooth_1}, the desired inequality is equivalent to 
\begin{align}\label{ineq22}
\frac{\log(\tilde{s})}{\tilde{s}^{\alpha}} \le  \frac{\log(\tilde{a})}{\tilde{a}^{\alpha}}.
\end{align}
Differentiating the left-hand side, we get \begin{align*}
\left(\frac{\log(\tilde{s})}{\tilde{s}^{\alpha}}\right)' = \frac{\tilde{s}^{\alpha - 1} - \alpha \tilde{s}^{\alpha - 1}\log(\tilde{s})}{\tilde{s}^{2\alpha}} = \frac{1 - \alpha\log(\tilde{s})}{\tilde{s}^{\alpha+1}}.
\end{align*}
The  function $\frac{\log(\tilde{s})}{\tilde{s}^{\alpha}}$ is strictly increasing for $\tilde{s} <\exp({\alpha^{-1}}).$ Hence, inequality  \eqref{ineq22} follows. 

\subsection{XOU Model}

\paragraph*{Proof of Theorem \ref{thm:XOU_exp_V}}\label{{pf:XOU_exp_V}} \textbf{(Exponential Utility).}
 Recall that the functions  $F$ and $G$ (see \eqref{F} and \eqref{G}) are respectively increasing and decreasing. Since   the exponential utility $U_e$ is strictly increasing, we postulate that the solution to the variational inequality  \eqref{xou_variational_inequality} is of the form $KF(z)$, where $K$ is a positive coefficient to be determined. By grouping $\nu$ with $\gamma$, the problem of selling $\nu$ units of the risky asset can be reduced to that of selling a single unit.   The   value-matching and  smooth-pasting conditions are
\begin{align}
KF(b_e) &= 1 - \exp\left(-\gamma \nu e^{b_e}\right),\label{eqn:XOUpasting_1}\\
KF'(b_e) &= \gamma \nu e^{b_e} \exp\left(-\gamma \nu e^{b_e}\right).  \label{eqn:XOUpasting_2}
\end{align}
Using   \eqref{eqn:XOUpasting_1}, we have $K =  {(1 - \exp\left(-\gamma \nu e^{b_e}\right))}{F(b_e)^{-1}} > 0.$ Combining \eqref{eqn:XOUpasting_1} and \eqref{eqn:XOUpasting_2}, we obtain \eqref{VXOU_exp_eqn} for $b_e$.

%
%Using the definitions of $F$ and $G$, we have the limits
%\begin{align*}
%\lim_{z \rightarrow -\infty}\frac{1-\exp\left(-\gamma \nu e^z\right)}{G(z)} = \lim_{z \rightarrow +\infty}\frac{1-\exp\left(-\gamma \nu e^z\right)}{F(z)} = 0.
%\end{align*}

Next, we want to establish that 
\begin{align}\label{xouexpbounded}
\E_z\left\{\int_{0}^{\infty} e^{-rt} \left\lvert (\L^Z-r)\left(1 - \exp\left(-\gamma \nu e^{Z_t}\right)\right)\right\rvert \,dt \right\} < \infty.
\end{align}
First, using \eqref{genOU} we compute  
\begin{align*}
(\L^Z-r)\left(1 - \exp\left(-\gamma \nu e^z\right)\right) &= \frac{\eta^2}{2}\gamma \nu e^z \exp\left(-\gamma \nu e^z\right)\left( 1 - \gamma \nu e^z\right) + \kappa(\theta - z)\gamma \nu e^z \exp\left(-\gamma \nu e^z\right) \\
&\quad- r\left(1-\exp\left(-\gamma \nu e^z\right)\right).
\end{align*}
Then, for any $T > 0,$
\begin{align*}
 \E_z&\left\{\int_{0}^{T} e^{-rt} \left\lvert (\L^Z-r)\left(1 - \exp\left(-\gamma \nu e^{Z_t}\right)\right)\right\rvert \,dt \right\} \notag\\
&<  \E_z\left\{\int_{0}^{T} e^{-rt} \left(\frac{\eta^2}{2}\left( 1 + \gamma \nu e^{Z_t}\right) + \kappa|\theta| + \kappa |Z_t| + r\right) \,dt \right\}\notag\\
&= \int_{0}^{T} e^{-rt}\left( \frac{\eta^2}{2} + \frac{\eta^2 \gamma \nu}{2} \E_z\left\{e^{Z_t}\right\} + \kappa|\theta| + \kappa \E_z\left\{|Z_t|\right\} + r \right) \,dt\notag\\
&< \frac{\eta^2}{2r} + \frac{\eta^2 \gamma \nu}{2r} e^{|z| + |\theta| + \frac{\eta^2}{4\kappa}} + \frac{\kappa|\theta|}{r} + \frac{\kappa}{r} \left(\sqrt{\frac{\eta^2}{\pi \kappa}} + |z| + |\theta| \right) + 1,
\end{align*}
where we have used the fact that $|Z_t|$ conditioned on $Z_0 = z$ has a folded normal distrbution. Furthermore, since this bound is time-independent and finite, we deduce that \eqref{xouexpbounded} is indeed true.
We follow the arguments in Section 2 of \cite{zervos2011buy} to   obtain the representation  
\begin{align}\label{xou_exp_rep}
1 - \exp\left(-\gamma \nu e^z\right) = &-G(z)\int_{-\infty}^z \Psi^Z(\upsilon)(\L^{Z}-r)(1 - \exp\left(-\gamma \nu e^{\upsilon}\right))\dx{\upsilon} \notag \\
& -F(z)\int_{z}^{+\infty} \Phi^Z(\upsilon)(\L^{Z}-r)(1 - \exp\left(-\gamma \nu e^{\upsilon}\right))\dx{\upsilon},
\end{align}
where
\begin{align}\label{psi_phi}
\Psi^Z(z) := \frac{2F(z)}{\eta^2\W^Z(z)}, \quad \quad \Phi^Z(z) := \frac{2G(z)}{\eta^2\W^Z(z)}, 
\end{align}
and
\begin{align*}
\W^Z(z) = F'(z)G(z)-F(z)G'(z) >0, \quad \forall z \in \R.
\end{align*}
To see the connection between \eqref{xou_exp_rep} and \eqref{VXOU_exp_eqn}, we divide both sides of \eqref{xou_exp_rep} by $F(z)$ and differentiate with respect to $z$, and get the derivative 
\begin{align}
\left(\frac{1 - \exp\left(-\gamma \nu e^z\right)}{F(z)}\right)' &=  \frac{\gamma \nu  e^{z} \exp\left(-\gamma \nu   e^{z}\right)F(z) - \left(1 - \exp\left(-\gamma \nu  e^{z}\right)\right)F'(z)}{F^2(z)} \label{xou_exp_derivative}\\
&= -\left(\frac{G(z)}{F(z)}\right)'\int_{-\infty}^z \Psi^Z(\upsilon)(\L^Z-r)\left(1 - e^{-\gamma \nu e^\upsilon}\right)\dx{\upsilon}\notag\\
&\quad \, -\frac{G(z)}{F(z)}\Psi^Z(z)(\L^Z-r)\left(1 - e^{-\gamma \nu e^z}\right)\notag \\
&\quad \, + \Phi^Z(z)(\L^Z-r)\left(1 - e^{-\gamma \nu e^z}\right)\notag\\
&= \frac{\W^Z(z)}{F^2(z)}\int_{-\infty}^z \Psi^Z(\upsilon)(\L^Z-r)\left(1 - e^{-\gamma \nu e^\upsilon}\right)\dx{\upsilon} = \frac{\W^Z(z)}{F^2(z)}\tilde{q}_e(z),\label{endeqnwfq}
\end{align}
where
\begin{align}\label{q_xou_exp}
\tilde{q}_e(z) := \int_{-\infty}^z \Psi^Z(\upsilon)(\L^Z-r)\left(1 - \exp\left(-\gamma \nu e^{\upsilon}\right)\right)\dx{\upsilon}.\end{align}
By comparing \eqref{VXOU_exp_eqn} to  the numerator on the RHS of \eqref{xou_exp_derivative}, and given that  $\frac{\W^Z(z)}{F^2(z)}$ in  \eqref{endeqnwfq} is strictly positive, we see that   the equation satisfied by $b_e$  in \eqref{VXOU_exp_eqn} is equivalent to $\tilde{q}_e(b_e) = 0.$   Therefore, our goal is   to show that $\tilde{q}_e(z)$ has a unique and finite root. 

Differentiating \eqref{q_xou_exp} with respect to $z$ yields
\begin{align*}
\tilde{q}_e'(z) = \Psi^Z(z)(\L^Z-r)\left(1 - \exp\left(-\gamma \nu e^z\right)\right).
\end{align*}
Furthermore, observe that the sign of $\tilde{q}_e'$ depends solely on $(\L^Z-r)\left(1 - \exp\left(-\gamma \nu e^z\right)\right).$ 
%Using \eqref{genOU}, we obtain
%\begin{align*}
%(\L^Z-r)\left(\exp\left(-\gamma \nu e^z\right)\right) &= \frac{\eta^2}{2}\gamma \nu e^z \exp\left(-\gamma \nu e^z\right)\left( 1 - \gamma \nu e^z\right) + \kappa(\theta - z)\gamma \nu e^z \exp\left(-\gamma \nu e^z\right) - r\left(\exp\left(-\gamma \nu e^z\right)\right).
%\end{align*}
We proceed to show that  $\tilde{q}_e'(z)$ has a unique root. To facilitate computation, we   define a new function
\begin{align*}
h(z) := \frac{\tilde{q}_e'(z)}{\Psi^Z(z)}\times \frac{\exp\left(\gamma \nu e^z\right)}{\gamma \nu e^z} &= (\L^Z-r)\left(1 - \exp\left(-\gamma \nu e^z\right)\right)\frac{\exp\left(\gamma \nu e^z\right)}{\gamma \nu e^z}\\
&\,= \frac{\eta^2}{2}\left( 1 - \gamma \nu e^z\right) + \kappa(\theta - z) - r\left(\frac{\exp\left(\gamma \nu e^z\right)}{\gamma \nu e^z} - \frac{1}{\gamma \nu e^z}\right).
\end{align*}
Since $h$ is obtained through dividing and multiplying $\tilde{q}_e'$ by strictly positive terms, any root of $\tilde{q}_e'$ must also be a root of $h$ and vice-versa.

To find the root of  $h$, we   solve  
\begin{align}\label{LUlevel}
-\frac{\eta^2\gamma \nu}{2}e^z - \frac{r}{\gamma \nu}e^{-z}\left(\exp\left(\gamma \nu e^z\right) - 1\right) = \kappa z - \kappa \theta - \frac{\eta^2\gamma \nu}{2}.
\end{align}
The RHS of \eqref{LUlevel} is a strictly increasing linear function. As for the LHS, we observe that 
\begin{align*}
\lim_{z \rightarrow +\infty} -\frac{\eta^2\gamma \nu}{2}e^z - \frac{r}{\gamma \nu}e^{-z}\left(\exp\left(\gamma \nu e^z\right) - 1\right) &= -\infty,\\
\lim_{z \rightarrow -\infty} -\frac{\eta^2\gamma \nu}{2}e^z - \frac{r}{\gamma \nu}e^{-z}\left(\exp\left(\gamma \nu e^z\right) - 1\right) &= r.
\end{align*}
Hence, in order for $h$ to have a unique root, it suffices to show that the LHS of \eqref{LUlevel} is strictly decreasing. Given that   $r$, $\gamma$, $\nu > 0 $ and $e^z$ is   strictly increasing, it   remains to show that the function $e^{-z}\left(\exp\left(\gamma \nu e^z\right) - 1\right)$ is strictly increasing for all $z \in \R$. The quotient rule gives \begin{align*}
\left(\frac{\exp\left(\gamma \nu e^z\right) - 1}{e^z} \right)' &= \frac{\exp\left(\gamma \nu e^z\right)\left(\gamma \nu e^z -1\right) + 1}{e^z}.
\end{align*}
The numerator $\exp\left(\gamma \nu e^z\right)\left(\gamma \nu e^z -1\right) + 1$ goes to $+\infty$ as $z$ goes to $+\infty$ and goes to $0$ as $z$ goes to $-\infty.$ Moreover, the derivative of $\exp\left(\gamma \nu e^z\right)\left(\gamma \nu e^z -1\right) + 1$ is $\gamma^2 \nu^2 e^{2z} \exp\left(\gamma \nu e^z\right)$ which is strictly positive. This proves that the function $e^{-z}\left(\exp\left(\gamma \nu e^z\right) - 1\right)$ is indeed strictly increasing and as a result, $h$ has a unique root, denoted  by $\zeta$. Finally, observe that 
\begin{align}\label{qsign}
\tilde{q}_e'(z) = \begin{cases}
> 0 &\, \textrm{ if }\, z < \zeta,\\
< 0 &\, \textrm{ if }\, z > \zeta.
\end{cases}
\end{align}
Combining \eqref{qsign}  with $\lim_{z\to-\infty}\limits \tilde{q}_e(z)=0$, we now see  that a unique root, $b_e$, such that $\tilde{q}_e(b_e)=0$, exists  if and only if $\lim_{z \to +\infty}\limits \tilde{q}_e(z) < 0$. To examine  this limit, we apply the definition of $F$ to get 
\begin{align}
\tilde{q}_e(z) &=  \frac{F^2(z)}{\W^Z(z)}\left(\frac{1 - \exp\left(-\gamma \nu e^z\right)}{F(z)}\right)',\label{derivative} \\ \quad \quad \frac{1 - \exp\left(-\gamma \nu e^z\right)}{F(z)} &> 0, \quad \forall z \in \R, \quad  \quad \lim_{z \to +\infty}\frac{1 - \exp\left(-\gamma \nu e^z\right)}{F(z)} = \frac{1}{+\infty} = 0.\label{derivative2}
 \end{align}
Since $\tilde{q}_e$ is strictly decreasing in $(\zeta, +\infty)$,   \eqref{derivative} and \eqref{derivative2} hold   if and only if $\lim_{z \to +\infty}\limits \tilde{q}_e(z) < 0$. This shows, as desired,  that there exists a unique and finite $b_e$  such that 
\begin{align*}
\gamma \nu e^{b_e} \exp\left(-\gamma \nu e^{b_e}\right)F(b_e) = \left(1 - \exp\left(-\gamma \nu e^{b_e}\right)\right)F'(b_e).
\end{align*} 
Moreover, by  \eqref{qsign}, we see that
\begin{align}\label{XOUprop}
b_e>\zeta \quad \textrm{and} \quad \tilde{q}_e(z) > 0, \quad \forall z < b_e.
\end{align}

Now, in order to ascertain the optimality of $\VV_e$ presented in \eqref{XOU_exp_V_sol}, we   re-express $\VV_e$ in terms of the variable $z$, and show that it  satisfies the following variational inequality:
\begin{align*}
\textrm{max}\{(\L^Z - r)\VV_e(e^z), \left(1 - \exp\left(-\gamma \nu e^z\right) \right) - \VV_e(e^z)\} = 0, \quad \forall z \in \R.
\end{align*}
Indeed, this follows from direct substitution. First,  on $[b_e, +\infty),$ we have 
\begin{align*}
\left(1 - \exp\left(-\gamma \nu e^z\right) \right) - \VV_e(e^z) = \left(1 - \exp\left(-\gamma \nu e^z\right) \right) - \left(1 - \exp\left(-\gamma \nu e^z\right) \right) = 0.
\end{align*}
On     $(-\infty,b_e)$, we apply \eqref{derivative} and \eqref{XOUprop} to conclude that  
\begin{align*}
\left(1 - \exp\left(-\gamma \nu e^z\right) \right) - \VV_e(e^z) = \left(1 - \exp\left(-\gamma \nu e^z\right) \right) - \frac{1 - \exp\left(-\gamma \nu e^{b_e}\right)}{F(b_e)}F(z) \le 0.
\end{align*} 
%Using  , we  follows from \begin{align*}
%\frac{1 - \exp\left(-\gamma \nu e^z\right)}{F(z)} \leq \frac{1 - \exp\left(-\gamma \nu e^{b_e}\right)}{F(b_e)}, \quad \forall z < b_e,
%\end{align*}

   Next, we   verify that $(\L^Z - r)\VV_e(e^z) \leq 0,$ $\forall z \in \R$. To this end, we have 
\begin{align*}
(\L^Z - r)\VV_e(e^z) &= (\L^Z - r)KF(z) = 0, \quad \text{ on } (-\infty, b_e),\\
(\L^Z - r)\VV_e(e^z) &= (\L^Z - r)\left(1 - \exp\left(-\gamma \nu e^z\right) \right) \leq 0, \quad \text{ on } [b_e, +\infty),
\end{align*}
as a consequence  of \eqref{LUXOU} and  \eqref{qsign}. Hence, we conclude  the optimality of $\VV_e$ in \eqref{XOU_exp_V_sol}.

\paragraph*{Proof of Theorem \ref{thm:XOU_log_V}}\label{pf:XOU_log_V} \textbf{(Log Utility).}
Since   $\log(\nu X) = Z + \log(\nu)$ where $Z$ is  an OU process,   the optimal asset sale problem can be viewed as  that under an OU process with a linear utility and a transaction cost (resp. reward) of value $\log(\nu)$ if $\nu <1$ (resp. $\nu >1$) (see \cite{LeungLi2014OU}).

The functions $F$ and $G$ given in \eqref{F} and \eqref{G} are respectively strictly increasing and decreasing functions. For any given $\nu,$ $Z + \log(\nu)$ is also a strictly increasing function. This prompts us to postulate a solution to the variational inequality \eqref{xou_variational_inequality} of the form $DF(z)$ where $D>0$ is a constant to be determined. Consequently, the optimal log-price thershold $b_l$ is determined from the following value-matching and smooth-pasting conditions: 
\begin{align}\label{xou_log_smoothpasting}
DF(b_l) = b_l + \log(\nu),\quad \text{ and } \quad 
DF'(b_l) = 1.
\end{align}
Combining these equations leads to \eqref{coefD} and  \eqref{VXOU_log_eqn}.
%Using the definitions of $F$ and $G$, we have
%\begin{align*}
%\lim_{z\rightarrow -\infty}\frac{z + \log(\nu)}{G(z)} = \lim_{z\rightarrow -\infty}\frac{1}{G'(z)} = 0, \quad \textrm{ and } \quad \lim_{z \rightarrow +\infty}\frac{z + \log(\nu)}{F(z)} = \lim_{z \rightarrow +\infty}\frac{1}{F'(z)} = 0.
%\end{align*}
Straightforward computation yields
\begin{align*}
(\L^{Z}-r)(z + \log(\nu)) &= -(\kappa + r)z + \kappa\theta - r \log(\nu),
\end{align*}
which is a strictly  decreasing linear function with a unique root $\ell.$ 
For any $T > 0$,
\begin{align*}
\E_{z}\left\{\int_{0}^{T} e^{-rt} \left\lvert (\L^{Z}-r)(Z_t + \log(\nu))\right\rvert \,dt \right\} &< \E_{z}\left\{\int_{0}^{T} e^{-rt} \left( (\kappa + r)|Z_t| + \kappa|\theta| + r|\log(\nu)|\right)\,dt \right\}\\
&= \int_{0}^{T} e^{-rt} \left( (\kappa + r)\E_{z}\left\{|Z_t|\right\} + \kappa|\theta| + r|\log(\nu)|\right)\,dt \\
&< \frac{\kappa + r}{r} \left(\sqrt{\frac{\eta^2}{\pi \kappa}} + |z| + |\theta| \right) + \frac{\kappa|\theta|}{r} + |\log(\nu)|, 
\end{align*}
which implies that
\begin{align*}
\E_{z}\left\{\int_{0}^{\infty} e^{-rt} \left\lvert (\L^{Z}-r)(Z_t + \log(\nu))\right\rvert \,dt \right\} < \infty.
\end{align*}
With this, we follow  the arguments in Section 2 of \cite{zervos2011buy} to obtain the   representation\begin{align}\label{xou_log_rep}
z + \log(\nu) = &-G(z)\int_{-\infty}^{z} \Psi^Z(\upsilon)(\L^{Z}-r)(\upsilon + \log(\nu))\dx{\upsilon} 
-F(z)\int_{z}^{+\infty} \Phi^Z(\upsilon)(\L^{Z}-r)(\upsilon + \log(\nu))\dx{\upsilon},
\end{align}
where $\Psi^Z$ and $\Phi^Z$ are as defined in \eqref{psi_phi}. 
We   relate \eqref{xou_log_rep} to \eqref{VXOU_log_eqn} by first dividing both sides of \eqref{xou_log_rep} by $F(z)$ and differentiating with respect to $z.$ This yields
\begin{align}
\left(\frac{z + \log(\nu)}{F(z)}\right)'&=  \frac{F(z) - F'(z)(z + \log(\nu))}{F^2(z)} \label{xou_log_derivative}\\
&=  -\left(\frac{G(z)}{F(z)}\right)'\int_{-\infty}^{z} \Psi^{Z}(\upsilon)(\L^{Z}-r)(\upsilon + \log(\nu))\dx{\upsilon} \notag\\
& \quad -\frac{G(z)}{F(z)} \Psi^{Z}(z)(\L^{Z}-r)(z + \log(\nu)) + \Phi^{Z}(z)(\L^{Z}-r)(z + \log(\nu)) \notag \\
&= \frac{\W^Z(z)}{F^2(z)}\int_{-\infty}^{z} \Psi^Z(\upsilon)(\L^{Z}-r)(\upsilon + \log(\nu)) \dx{\upsilon} = \frac{\W^Z(z)}{F^2(z)}\tilde{q}_l(z), \notag
\end{align}
where
\begin{align*}
\tilde{q}_l(z) := \int_{-\infty}^{z} \Psi^Z(\upsilon)(\L^{Z}-r)(\upsilon + \log(\nu))\dx{\upsilon}.
\end{align*}
Comparing   \eqref{VXOU_log_eqn} and the RHS of  \eqref{xou_log_derivative}, along with the facts that  $F > 0$ and $\W^Z > 0$, we see that solving  equation  \eqref{VXOU_log_eqn} for the log-price threshold is equivalent to solving 
$$
\tilde{q}_l(z)=0.
$$
Direct differentiation yields that 
%\begin{align*}
%\tilde{q}_l'(z) = \Psi^Z(z)(\L^{Z}-r)(z + \log(\nu)).
%\end{align*}
%%Straightforward computation yields
%%\begin{align*}
%%(\L^{Z}-r)(z + \log(\nu)) &= -(\kappa + r)z + \kappa\theta - r \log(\nu),
%%\end{align*}
%%which is a straightly decreasing linear function with a unique root $\ell.$ 
%As a consequence, 
\begin{align}\label{qsign_log}
\tilde{q}_l'(z) = \Psi^Z(z)(\L^{Z}-r)(z + \log(\nu)) \,\,\begin{cases}
> 0 &\, \textrm{ if }\, z < \ell,\\
< 0 &\, \textrm{ if }\, z > \ell.
\end{cases}
\end{align}

The fact that $\lim_{z\to-\infty}\tilde{q}_l(z)=0$ implies that there exists a unique $b_l$ such that $\tilde{q}_l(b_l)=0$ if and only if $\lim_{z\to +\infty}\tilde{q}_l(z) < 0$. By the definition of $F$, we have
\begin{align}\label{derivative_log}
\tilde{q}_l(z) = \frac{F^2(z)}{\W^Z(z)}\left(\frac{z + \log(\nu)}{F(z)}\right)', \quad \quad\frac{z + \log(\nu)}{F(z)} > 0, \quad \forall z > -\log(\nu), \quad  \quad \lim_{z \to +\infty}\frac{z + \log(\nu)}{F(z)} = 0. 
\end{align}
Given that $\tilde{q}_l$ is strictly decreasing in $(\ell, +\infty),$ we conclude that in order for \eqref{derivative_log} to hold, we must have $\lim_{z \to +\infty}\tilde{q}_l(z) < 0$. This means that there exists a unique  and finite $b_l$  such that $(b_l + \log(\nu)) F'(b_l) = F(b_l)$. Moreover, given \eqref{qsign_log}, we have
\begin{align}\label{XOUprop_log}
b_l>\ell \quad \textrm{and} \quad \tilde{q}_l(z) > 0, \quad \forall z < b_l.
\end{align}
Furthermore, since both $F$ and $F'$ are strictly positive, $b_l + \log(\nu)$ must also be positive.

Lastly, we need to check that the following variational inequality holds for any fixed $\nu$:
\begin{align}\label{XOU_log_V_variational_inequality}
\textrm{max}\{(\L^{Z} - r)\VV_l(e^z), (z + \log(\nu)) - \VV_l(e^{z})\} = 0, \quad \forall z \in \R.
\end{align}
To begin, on the interval $[b_l, +\infty),$ we have $(z + \log(\nu)) - \VV_l(e^{z}) = (z + \log(\nu)) - (z + \log(\nu)) = 0.$
Next, on the region $(-\infty,b_l)$,  we have
\begin{align*}
(z + \log(\nu)) - \VV_l(e^{z}) = (z + \log(\nu)) - \frac{b_l +  \log(\nu)}{F(b_l)}F(z) \leq 0 
\end{align*}
  since the function $\frac{z + \log(\nu)}{F(z)}$ is increasing on $(-\infty,b_l)$ due to \eqref{derivative_log} and \eqref{XOUprop_log}. Also, we note that
\begin{align*}
(\L^{Z} - r)\VV_l(e^z) = (\L^{Z} - r)DF(z) &= 0,  \quad \text{ on } (-\infty, b_l), \\
(\L^{Z} - r)\VV_l(e^{z}) = (\L^{Z}- r)(z + \log(\nu))&\leq 0, \quad \text{ on } [b_l, +\infty). 
\end{align*}
The latter inequality is true due to \eqref{qsign_log} and \eqref{XOUprop_log}. $\VV_l$ defined in Theorem \ref{thm:XOU_log_V} is therefore the optimal solution to the optimal asset sale problem under log utility.

\paragraph*{Proof of Theorem \ref{thm:XOU_pow_V}}\label{pf:XOU_pow_V}\textbf{(Power Utility).}  Since the powered XOU process, $X^p$, is still an XOU process, the asset sale problem is an optimal stopping problem driven an XOU process, which  has been solved by the authors' prior work; see Theorem 3.1.1 of \cite{LeungLiWang2014XOU}. Therefore, we omit the proof. 

\bibliographystyle{apa}
\linespread{-0.4}
%\singlespacing
\begin{small}
\bibliography{mybib2_10222015}
\end{small}

\end{document}